\documentclass[
reprint,  
 amsmath,
 amssymb,
 aps,
 prc,
]{revtex4-2}

\usepackage{graphicx}
\usepackage{tabularx}
\usepackage{dcolumn}

\usepackage{bm}
\usepackage{color}
\usepackage{physics}
\usepackage{url}
\usepackage{hyperref}
\hypersetup{
setpagesize=false,
 bookmarksnumbered=true,%
 bookmarksopen=true,%
 colorlinks=true,%
 linkcolor=blue,
 citecolor=blue,
}

\makeatletter
\newcommand{\colorcaption}[2][]{%
  \begingroup%
  \renewcommand{\@caption@fignum@sep}{ (color online).  }%
  \caption[#1]{#2}%
  \endgroup%
}
\makeatother

\begin{document}


\title{Directed flow in relativistic resistive magneto-hydrodynamic expansion for symmetric and asymmetric collision systems}

\author{Kouki Nakamura$^{1,2}$}
\email{knakamura@hken.phys.nagoya-u.ac.jp}
\author{Takahiro Miyoshi$^{2}$}%
\email{miyoshi@sci.hiroshima-u.ac.jp}
\author{Chiho Nonaka$^{1,2,3}$}%
 \email{nchiho@hiroshima-u.ac.jp}
\author{Hiroyuki R. Takahashi$^{4}$}%
 \email{takhshhr@komazawa-u.ac.jp}
\affiliation{%
$^1$Department of Physics, Nagoya University, Nagoya 464-8602, Japan \\
$^2$Department of Physics, Hiroshima University, Higashihiroshima 739-8526, Japan\\
$^3$Kobayashi Maskawa Institute, Nagoya University, Nagoya 464-8602, Japan\\
$^4$Department of Physics, Komazawa University, Tokyo 154-8525, Japan\\
 }%

\date{\today}

\begin{abstract}

  We construct a dynamical model for high-energy heavy-ion collision based on the relativistic resistive magneto-hydrodynamic framework.
  Using our newly developed (3+1)-dimensional relativistic resistive magneto-hydrodynamics code, we investigate magneto-hydrodynamic expansion in symmetric and asymmetric collision systems as a first application to high-energy heavy-ion collisions.
  As a realistic initial condition for electromagnetic fields, we consider the solutions of the Maxwell equations with the source term of point charged particles moving in the direction of the beam axis, including finite constant electrical conductivity of the medium.
  We evaluate the directed flow in the symmetric and asymmetric collisions at RHIC energy. 
  We find a significant effect of finite electrical conductivity on the directed flow in the asymmetric collision system.
  We confirm that a certain amount of energy transfer by dissipation associated with Ohmic conduction occurs in the asymmetric collision system because of asymmetry of the electric field produced by two different colliding nuclei. 
  Because this energy transfer makes the pressure gradient of the medium flatter, the growth of directed flow decreases.

\end{abstract}

\maketitle

\section{Introduction}\label{I}
One of the purposes of the high-energy heavy-ion collisions is an exploration of the phase diagram in quantum chromo-dynamics (QCD).
Since the strongly coupled quark-gluon plasma (QGP) was discovered at Relativistic Heavy Ion Collider (RHIC), a relativistic hydrodynamic model has been used as description of the space-time evolution of the hot and dense medium produced after the collisions~\cite{BACK200528,ADCOX2005184,ADAMS2005102,ARSENE20051}.
At the same time, the lower bound for the dimensionless ratio of shear viscosity to entropy density is evaluated to be $\eta/s = 1/4\pi$ by AdS/CFT correspondence~\cite{PhysRevLett.94.111601}.
This ratio takes the minimum around the critical temperature and can pinpoint the location of the QCD phase transition of rapid crossover from hadronic to QGP matter~\cite{PhysRevLett.97.152303}.
The situation triggered the construction and development of relativistic viscous hydrodynamic models~\cite{doi:10.1146/annurev-nucl-102212-170540,doi:10.1142/S0217751X13400113,DERRADIDESOUZA201635,PhysRevC.98.054906}. 
The relativistic hydrodynamic equation has close relation to the QGP bulk properties; the equation of state (EoS) and transport coefficients.
For the phase transition between the hadronic phase and the QGP phase, intensive studies are performed by lattice QCD and the parametrized EoS based on the analysis is now available~\cite{Borsanyi:2010cj,PhysRevD.85.054503,BLUHM2014157}.
On the other hand, there are no conclusive results for the transport coefficients.
In this situation, model-to-data comparison with Bayesian analysis plays an important role for evaluation of shear and bulk viscosities and the charge diffusion constant~\cite{Bernhard:2019bmu,PhysRevC.103.054904}.
Thus, the present target of the high-energy heavy-ion collisions is the quantitative study of QGP bulk properties, which is advanced from search for the existent of QGP.

In high-energy heavy-ion collisions, ultraintense electromagnetic fields are produced by the two colliding positively charged nuclei.
The intensity of the magnetic field in the event plane becomes large with increasing the center of mass energy, e.g., $|e\mathbf{B}| \sim 10^{14}$-$10^{15}~\mathrm{T} \sim m_\pi^{2}$ in $\sqrt{s_{\mathrm{NN}}} = 200~\mathrm{GeV}$ Au-Au collisions at RHIC~\cite{Huang_2016}.
Also, the electric field intensity is strong, because of large Lorentz factor of positively charged heavy nuclei in high-energy collisions.  
Such electromagnetic fields can affect the hydrodynamics of the created medium.
The effect of a strong magnetic field on the hydrodynamic evolution of the QGP medium has been studied based on a simplified form of the equation of the relativistic magneto-hydrodynamics (RMHD) such as reduced MHD~\cite{Roy:2017yvg} and relativistic ideal MHD with infinite electrical conductivity~\cite{Inghirami:2016iru,Inghirami:2019mkc}. 

The electrical conductivity characterizes the response of a medium to electromagnetic field.
Extracting the value of electrical conductivity from the experimental data at RHIC and the Large Hadron Collider (LHC) is an important subject for a detailed discussion of interesting phenomena under the high intense electromagnetic fields such as the chiral magnetic effect~\cite{KHARZEEV2008227,PhysRevD.78.074033} and vacuum birefringence of photon~\cite{PhysRevD.101.034015,PhysRevLett.127.052302}.
Temperature and external magnetic field dependencies on the electrical conductivity are investigated by the Lattice QCD~\cite{Aarts:2014nba,PhysRevD.102.054516}.
However, a dynamical model is needed for connection between the experimental data and the results of the first principle calculation.

Under some simplification, the effect of electromagnetic fields such as Lorentz force and Coulomb force on the relativistic viscous hydrodynamic expansion in the symmetric collision system has been discussed~\cite{Gursoy:2014aka,Pang:2016yuh,Roy:2017yvg,Gursoy:2018yai}. 
They found only small effect of electromagnetic fields to observables.
On the other hand, in the asymmetric collision system, a straightforward estimation of the electrical conductivity has been presented by focusing on the electric current~\cite{Hirono:2012rt}. 
In the context of the relativistic ideal magneto-hydrodynamic framework, the evolution of the electric field produced by colliding nuclei is neglected.
To handle the electric field produced by two different colliding nuclei, we need to construct the relativistic resistive magneto-hydrodynamic (RRMHD) model.
Hence, we consider the RRMHD framework in which Maxwell equations with finite electrical conductivity and relativistic hydrodynamic equations are simultaneously solved.
It is not possible to evaluate the electrical conductivity of the QGP medium from the analysis of high-energy heavy-ion collisions without using the RRMHD framework.

In this study, we construct a dynamical model for the high-energy heavy-ion collisions based on the RRMHD framework in the Milne coordinates and apply it to the symmetric and asymmetric collision systems.
Our model is built in a resistive extension of the relativistic ideal magneto-hydrodynamic model~\cite{Inghirami:2016iru,Inghirami:2019mkc}.
As a realistic initial condition of electromagnetic fields, we consider the solutions of Maxwell equations with the source term of the point charged particles moving in the direction of the beam axis and constant electrical conductivity of the medium~\cite{PhysRevC.88.024911}.
We evaluate the directed flow ($v_1$) in the symmetric and asymmetric high-energy heavy-ion collisions, using our RRMHD model. 

This paper is organized as follows.
In Sec.~\ref{II}, we briefly review the formulation and numerical models of the RRMHD systems in our simulation.
We apply our RRMHD simulation code to both of Au-Au and Cu-Au collisions in Sec.~\ref{IV}.
Numerical results are shown in Sec.~\ref{results} and a summary is given at the end in Sec.~\ref{V}.
Unless otherwise specified, we use natural units $\hbar = c = \epsilon_0 = \mu_0 = 1$, where $\epsilon_0$ and $\mu_0$ are the electric permittivity and the magnetic permeability in vacuum, respectively.
Throughout the paper the components of the four-tensors are indicated with greek indices, whereas three-vectors are denoted as boldface symbols. 

\section{Relativistic Resistive Magneto-Hydrodynamics}\label{II}

\subsection{formulations}
The RMHD framework is a model of interaction of conducting plasma and the electromagnetic fields~\cite{Goedbload, Anile}.
In this paper, we consider the RMHD framework with finite electrical conductivity with massless particles as the description of the space-time evolution of the coupled system of QGP with electromagnetic fields.
With the finite electrical conductivity of the plasma, the system follows the RRMHD equations constituted the conservation laws of fluid quantities and Maxwell equations~\cite{Komissarov:2007wk}.
The conservation laws for the charged current $N^\mu$ and for the total energy momentum tensor of the plasma $T^{\mu\nu}$ in the dynamics of whole system, are written by,
\begin{eqnarray}
  \label{ccc}
  \nabla_\mu N^\mu = 0,\\
  \label{emc}
  \nabla_\mu T^{\mu\nu} = 0,
\end{eqnarray}
where $\nabla_\mu$ is the covariant derivative.
The electro-magnetic fields follow Maxwell equations,
\begin{eqnarray}
  \label{maxwell1}
  \nabla_\mu F^{\mu\nu} = -J^{\nu},\\
  \label{maxwell2}
  \nabla_\mu ~^\star F^{\mu\nu} = 0,
\end{eqnarray}
where $F^{\mu\nu}$ is a Faraday tensor and $^{\star}F^{\mu\nu} = \frac{1}{2}\epsilon^{\mu\nu\rho\sigma}F_{\mu\nu}$ is it's dual tensor,
with $\epsilon^{\mu\nu\rho\sigma} = (-g)^{-1/2}[\mu\nu\rho\sigma]$, $g = \det(g_{\mu\nu}) $ and $[\mu\nu\rho\sigma]$ is a completely anti-symmetric tensor. 
Here we take the metric $\eta^{\mu\nu} = \rm{diag}(-1,1,1,1)$ in the Minkowski space-time.
If the magnetization and polarization effects are ignored, the energy-momentum tensor of the electromagnetic fields is known to be,
\begin{eqnarray}
  T^{\mu\nu}_{f} = F^{\mu\lambda}F_\lambda^\nu - \frac{1}{4}g^{\mu\nu}F^{\lambda\kappa}F_{\lambda\kappa},
\end{eqnarray}
and this tensor follows $\nabla_\mu T^{\mu\nu}_f = J_{\mu}F^{\mu\nu}$, from Maxwell equations.
The total energy momentum tensor is sum of the contribution of matter and electromagnetic fields $T^{\mu\nu} = T^{\mu\nu}_m + T^{\mu\nu}_f$.
The conservation law of the total system Eq.~(\ref{emc}) gives,
\begin{eqnarray}
  \nabla_\mu T^{\mu\nu}_m = -J_{\mu}F^{\mu\nu}.
\end{eqnarray}
In the ideal limit of the relativistic viscous hydrodynamics and the local equilibrium condition,
the energy momentum tensor and the charge current of fluids are written by,
\begin{eqnarray}
  N^\mu = \rho_B u^\mu,\\
  T^{\mu\nu}_m = (e+p)u^\mu u^\nu +pg^{\mu\nu},
\end{eqnarray}
where $u^{\mu}$ ($u^\mu u_\mu = -1$) is a single fluid four-velocity, $\rho_B$ is the baryon number density, $e = T^{\mu\nu}_m u_\mu u_\nu$ is energy density and $p =\frac{1}{3}\Delta_{\mu\nu}T^{\mu\nu}_m$ is pressure of the fluid.
We have introduced the projection tensor $\Delta_{\mu\nu}= g_{\mu\nu} + u_\mu u_\nu$.
The Faraday tensor and it's dual tensor are rewritten as,
\begin{eqnarray}
        F^{\mu\nu}= u^\mu e^\nu - u^\nu e^\mu + \epsilon^{\mu\nu\lambda\kappa}b_\lambda u_\kappa,\\
  ^\star F^{\mu\nu} = u^{\mu} b^{\nu} - u^{\nu} b^{\mu} - \epsilon^{\mu\nu\lambda\kappa}b_\lambda u_\kappa, 
\end{eqnarray}
where,
\begin{eqnarray}
  e^\mu = F^{\mu\nu}u_\nu,~( e^{\mu}u_\mu = 0 ),\\
  b^\mu = ~^\star F^{\mu\nu}u_\nu,~( b^{\mu}u_\mu = 0),
\end{eqnarray}
are the electric field and the magnetic field measured in the comoving frame of the fluid.
Since the system of equations~Eqs.(\ref{ccc})-(\ref{maxwell2}) is closed by the Ohm's law, we take into account the simplest form of it, including only plasma resistivity~\cite{Blackman:1993pbp}. In the covariant form, Ohm's law is written by,
\begin{equation}
  J^\mu = \sigma F^{\mu\nu}u^{\nu} + qu^\mu,
\end{equation}
where $\sigma$ is electrical conductivity and $q = -J^\mu u_\mu$ is electric charge density of the fluid in the comoving frame.
The presence of finite electrical conductivity in the plasma induces anisotropic magnetic dissipation and Joule heating. 
They affect the topology of the magnetic field line, which is known as magnetic reconnection discussed in some astrophysical applications~\cite{Takahashi_2011,Zenitani_2010,Watanabe_2006}.
The electrical conductivity plays an important role for the energy transfer from the electromagnetic fields to the fluid in RMHD.

\subsection{Numerical models}\label{III}
We now represent the equations of motion for the RRMHD in a suitable form for numerical calculation. 
We split the spacetime into 3 + 1 components by space-like hypersurface defined as the iso-surfaces of a scalar time function $t$ with a metric of the form,
 

\begin{equation}
  ds^2= -dx^0dx^0 + g_{ij}dx^idx^j.
\end{equation}
We introduce velocity $v^i$, electric fields $E^i$ and magnetic fields $B^i$ as measured in the laboratory frame.
The fluid four-velocity is rewritten as,
\begin{equation}
  u^\mu = (\gamma, \gamma v^i),
\end{equation}
where $\gamma = \sqrt{1 - v^iv_i}$ is the Lorentz factor of the fluid's flow.
We also define the electric and magnetic fields,
\begin{eqnarray}
  e^\mu = (\gamma v_kE^k, \gamma E^i + \gamma\epsilon^{ijk}v_jB_k),\\
  b^\mu = (\gamma v_kB^k, \gamma B^i - \gamma\epsilon^{ijk}v_jE_k). 
\end{eqnarray} 

Let us rewrite the equations of motion Eqs.~(\ref{ccc})-(\ref{maxwell2}) in a conservative form which is appropriate for numerical integration,
\begin{equation}\label{conservative form}
  \partial_0 \left(\sqrt{-g}~\mathbf{U}\right) + \partial_i \left(\sqrt{-g}~\mathbf{F}^i\right) = \sqrt{-g}~\mathbf{S},
\end{equation}
where $\mathbf{U}, \mathbf{F}^i$ and $\mathbf{S}$ are the set of conservative variables, numerical fluxes and source terms, respectively.
These variables contain the following components,
\begin{equation}
  \mathbf{U} = \mqty(\gamma \rho_B \\\Pi_j\\\varepsilon\\B^j\\E^j),
  \mathbf{F}^i = \mqty(\gamma\rho_Bv^i\\T^i_j\\\Pi^i\\\epsilon^{jik}E_k\\-\epsilon^{jik}B_k),
  \mathbf{S} = \mqty(0\\\frac{1}{2}T^{ik}\partial_j g_{ik}\\-\frac{1}{2}T^{ik}\partial_0g_{ik}\\0\\-J^i),
\end{equation}
where the total momentum $\Pi^i$, the stress tensor $T_{ij}$ and the total energy density $\varepsilon$ are given by,
\begin{eqnarray}
  \Pi_i &=& (e+p)\gamma^2v_i + \epsilon_{ijk}E^jB^k,\\
  T_{ij} &=& (e+p)\gamma^2v_iv_j + (p+p_{\rm{em}})g_{ij} - E_iE_j - B_iB_j,\nonumber\\
  \\
  \varepsilon &=& (e+p)\gamma^2 - p + p_{\rm{em}},
\end{eqnarray} 
where the electromagnetic energy density $p_{\mathrm{em}}$ is defined as $p_{\rm{em}} = \frac{1}{2}(E^2+B^2)$.
We assume that the fluid follows the ultrarelativistic ideal gas EoS, $e = p/3$.

In our new numerical code for RRMHD simulation, the time integration of the conservative variables is executed by the second order of Runge-Kutta algorithm~\cite{Komissarov:2007wk}.
The primitive variables are interpolated from cell center to cell surface by using the second order accurate scheme~\cite{VANLEER1977276}.
The constraints $\div B=0$ and $\div E = q$ should hold if they are satisfied at the initial state.
For numerical simulation, these conditions, however, sometimes are violated because of the numerical error, which leads to unphysical oscillation.
In this paper, we employ the generalized Lagrange multiplier method to guarantee these conditions~\cite{MUNZ2000484,Komissarov:2007wk,Porth:2016rfi}.
We note that timescales of the decay of electric field and diffusion of magnetic field are typically $1/\sigma$, which are sometimes much shorter than the dynamical timescale. To avoid the unexpected small time step in numerical simulation, we adopt the semi-analytic solutions to integrate Ampere's law~\cite{Komissarov:2007wk}.
We will show the details of our numerical algorithm and test problems for the verification of our numerical code in a paper to be published later.

\section{Application to High-Energy Heavy-Ion Collision}\label{IV}

We simulate the space-time evolution of the hot and dense medium with electromagnetic fields produced in high-energy heavy-ion collisions, utilizing the RRMHD framework in the Milne coordinates ($\tau,\bf{x}_{\rm{T}},\eta_s$), which are described by the specific time $\tau = \sqrt{t^2 - z^2}$, the coordinates in the transverse plane $\mathbf{x}_{\rm{T}} = (x,y)$, and the space rapidity $\eta_s = \frac{1}{2}\ln{\frac{t + z}{t - z}}$.

\subsection{Initial condition for the medium}

\begin{figure*}[t]
  \begin{minipage}[l]{0.45\linewidth}
    \includegraphics[width=8cm,height=6cm]{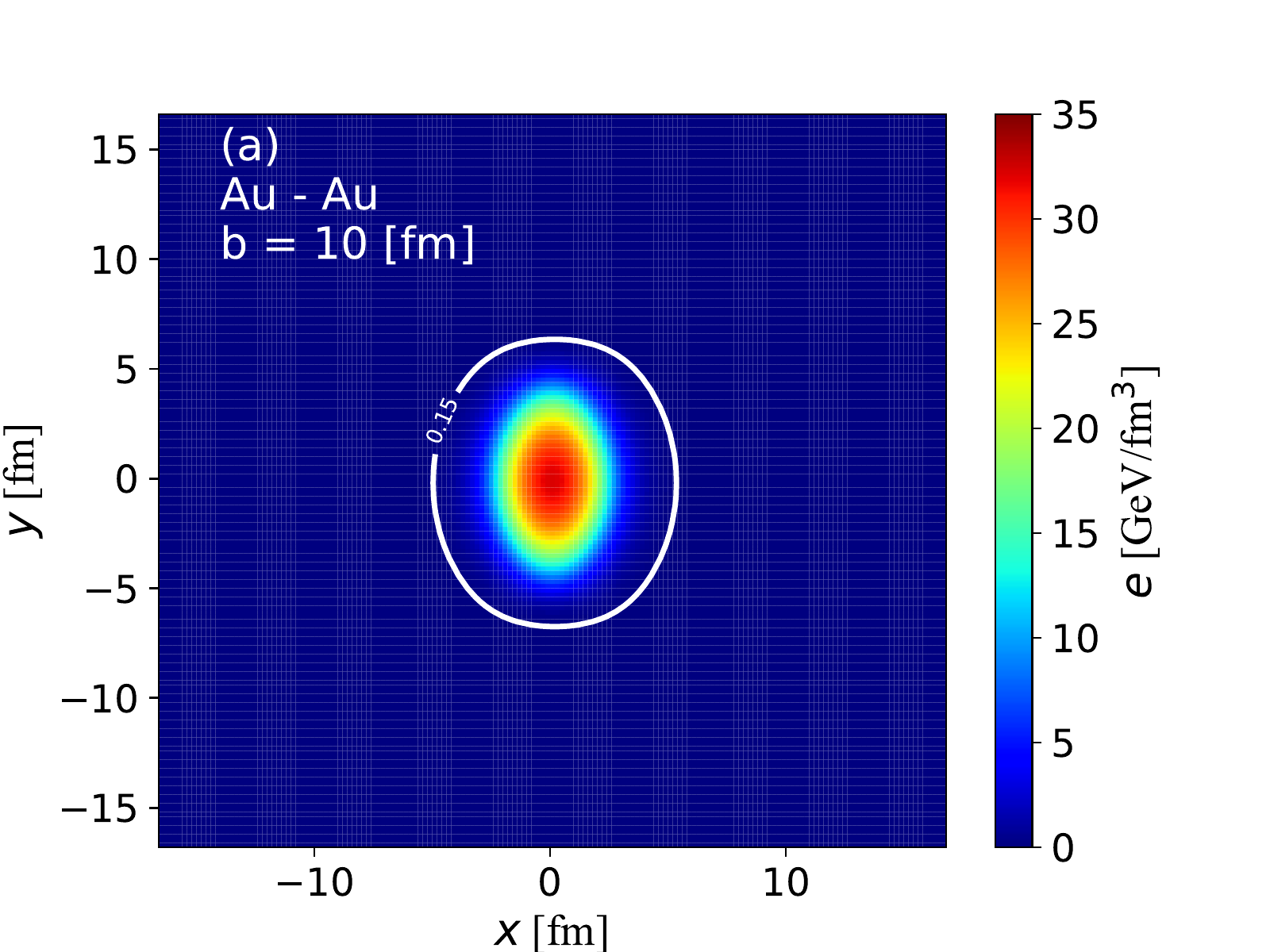}
  \end{minipage}
  \begin{minipage}[r]{0.45\linewidth}
    \includegraphics[width=8cm,height=6cm]{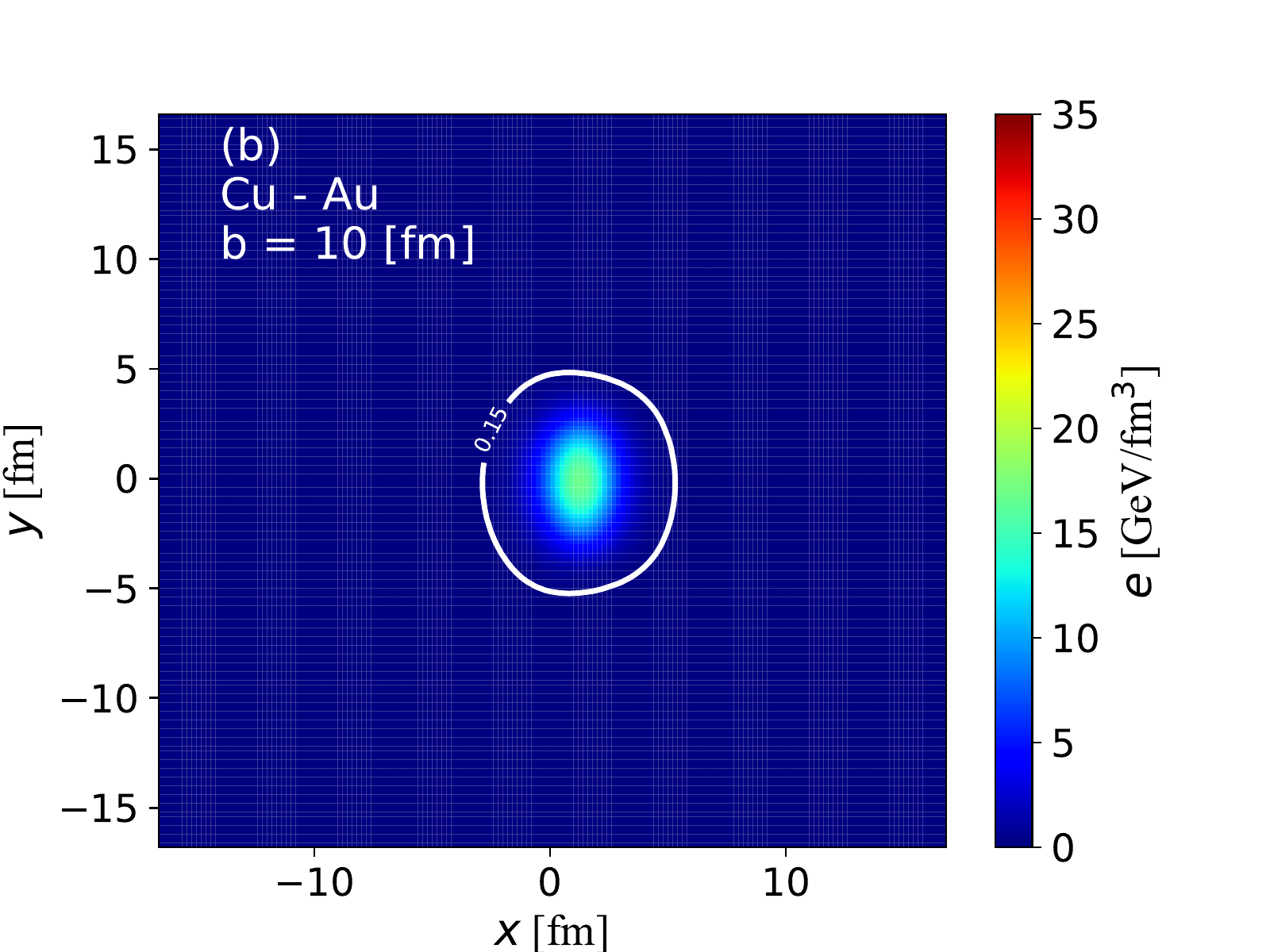}
     \end{minipage}
    \caption{(color online) The initial spatial distribution of the energy density in the transverse plane at $\eta_s = 0$. We display the cases of Au-Au collisions (a) and Cu-Au collisions (b), respectively.
    The white line represents the iso-thermal curve at $e = 0.15~\rm{GeV/fm^3}$.}
    \label{fig_ini:T}
\end{figure*}

\begin{figure*}[t]
  \begin{minipage}[l]{0.45\linewidth}
    \includegraphics[width=8cm,height=6cm]{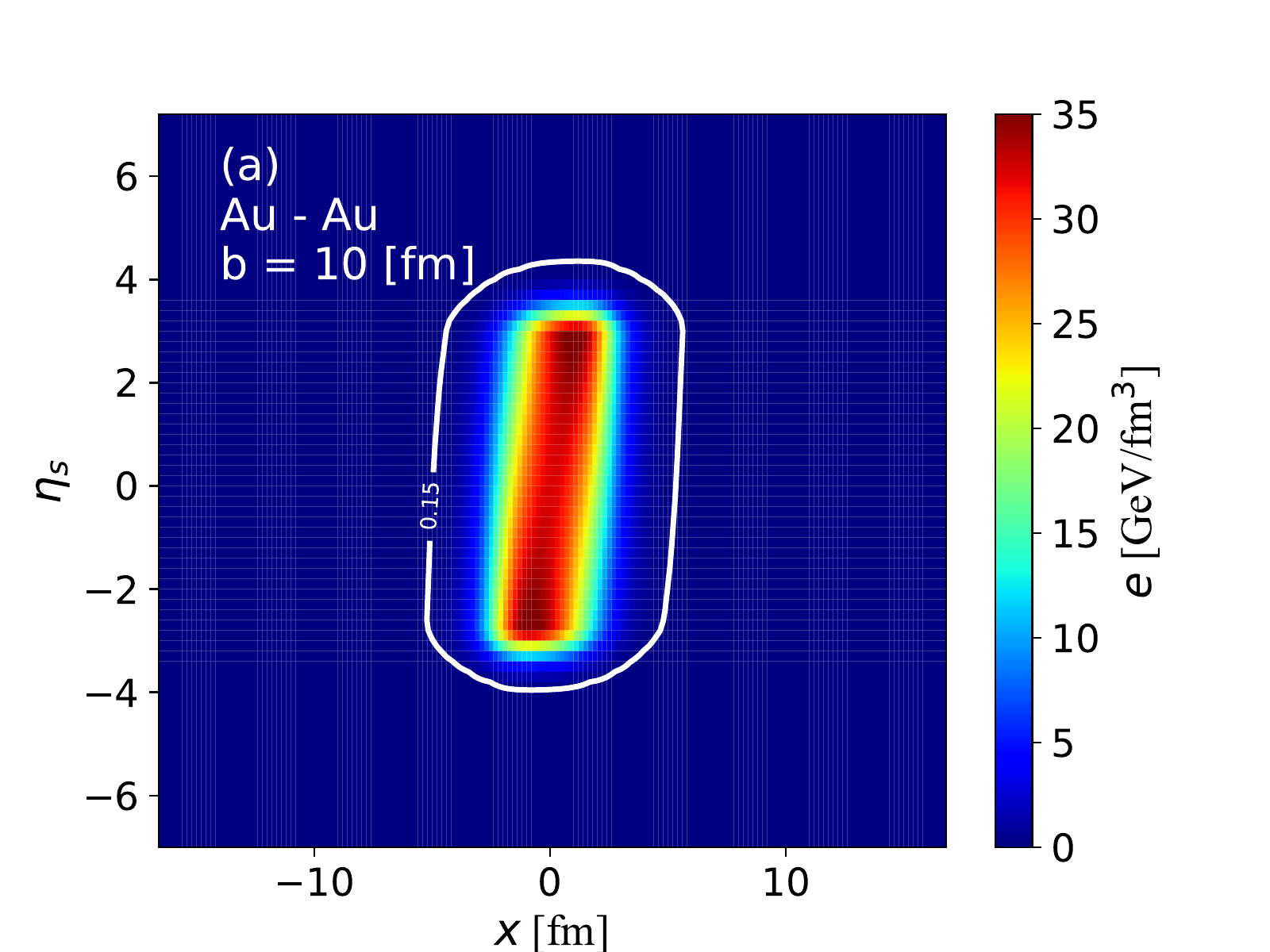}
  \end{minipage}
  \begin{minipage}[r]{0.45\linewidth}
    \includegraphics[width=8cm,height=6cm]{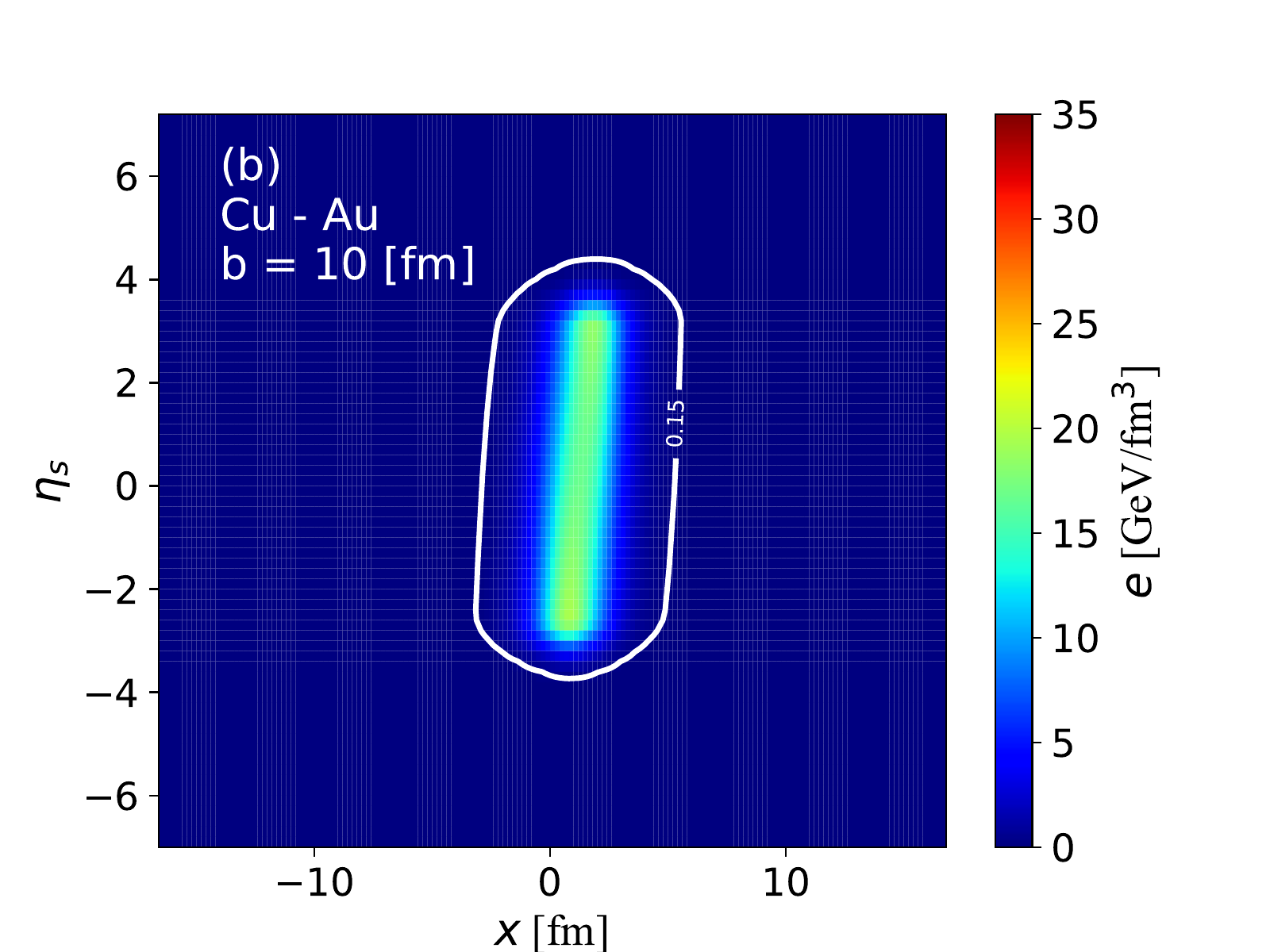}
     \end{minipage}
    \caption{(color online) The initial spatial distribution of the energy density in the reaction plane at $y = 0~\mathrm{fm}$. We show the cases of Au-Au collisions (a) and Cu-Au collisions (b), respectively. 
    The white line represents the iso-thermal curve at $e = 0.15~\rm{GeV/fm^3}$.
    }
    \label{fig_ini:L}
\end{figure*}

\begin{figure*}[t]
  \begin{minipage}[l]{0.45\linewidth}
    \includegraphics[width=8cm,height=6cm]{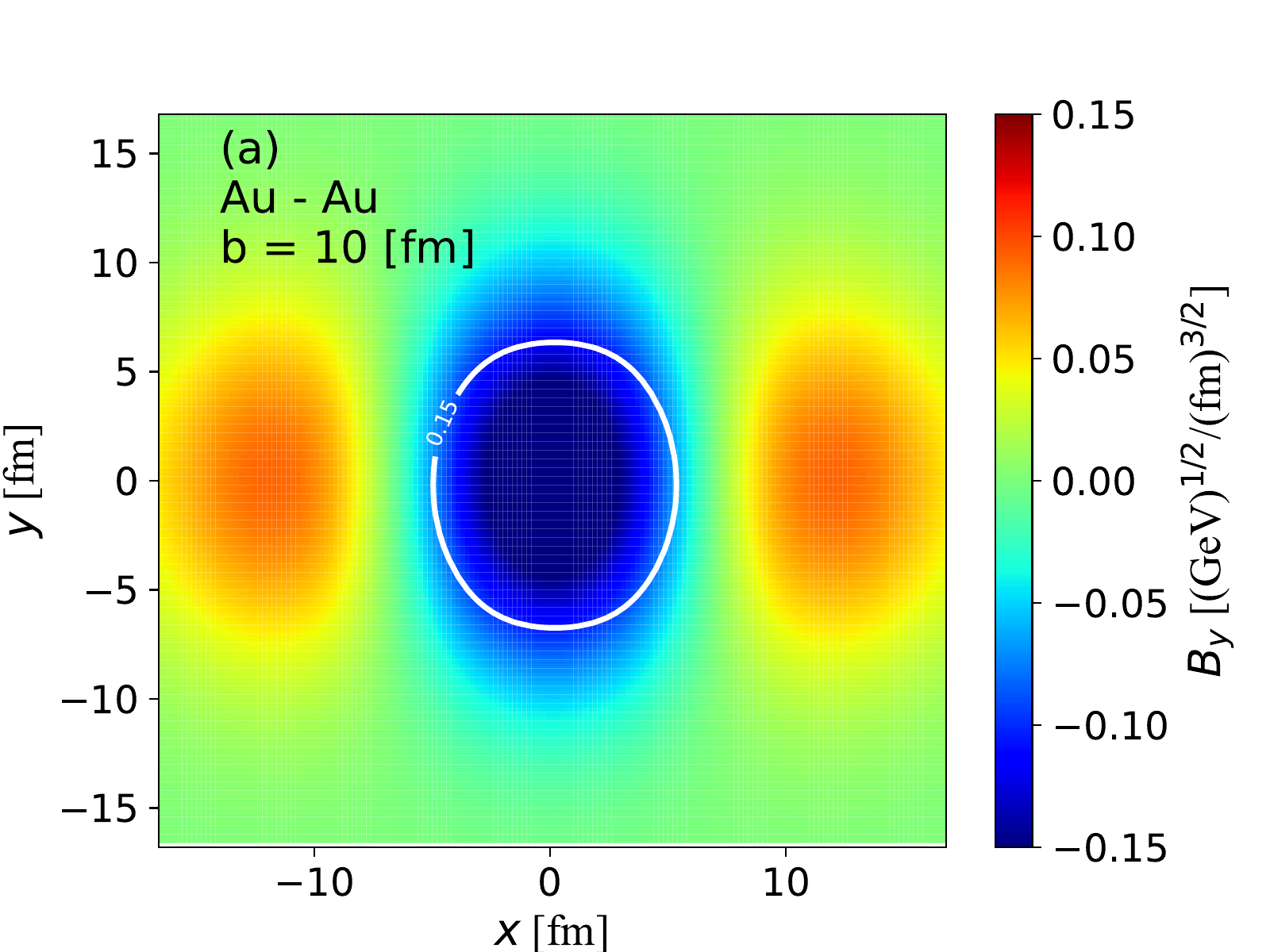}
  \end{minipage}
  \begin{minipage}[r]{0.45\linewidth}
    \includegraphics[width=8cm,height=6cm]{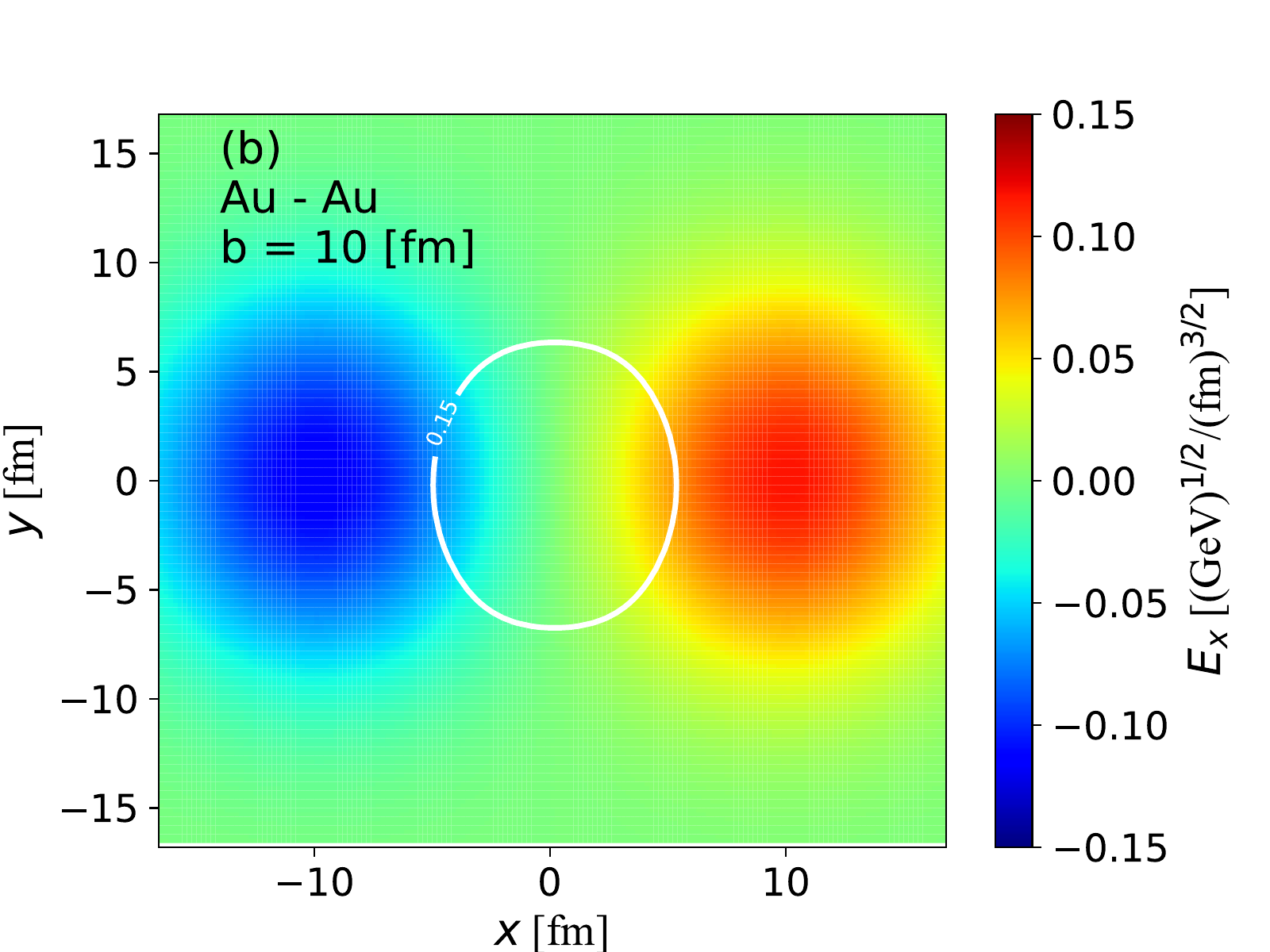}
     \end{minipage}
    \caption{(color online) The initial electromagnetic field in the transverse plane at $\eta_s=0$ for Au-Au collisions.
    We display the $y$-component of the magnetic field (a) and the $x$-component of the electric field (b), respectively.
    The white line represents the iso-thermal curve at $e = 0.15~\rm{GeV/fm^3}$.
    }
    \label{fig_ini:EM_AuAu_T}
\end{figure*}

\begin{figure*}[t]
  \begin{minipage}[l]{0.45\linewidth}
    \includegraphics[width=8cm,height=6cm]{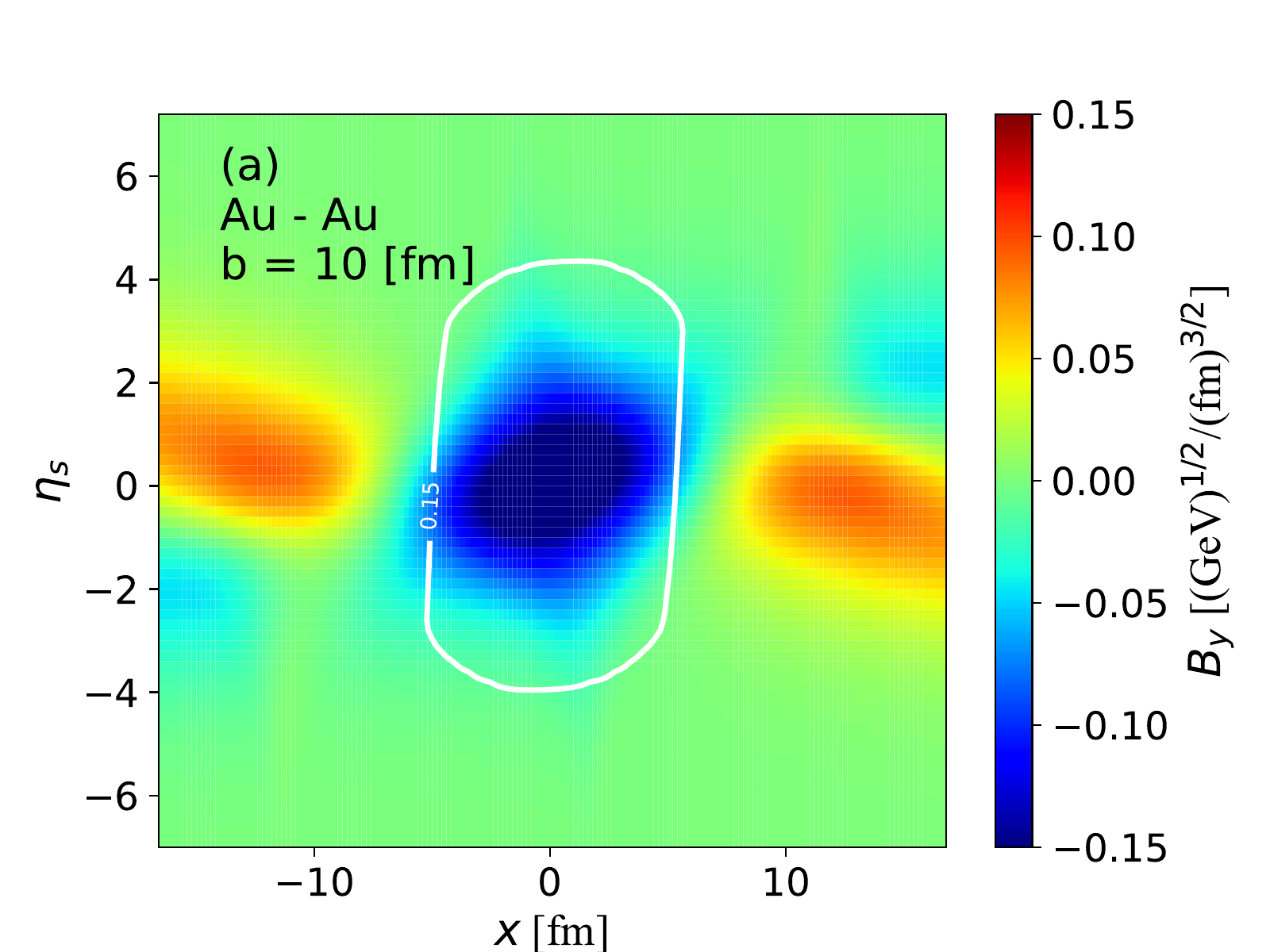}
  \end{minipage}
  \begin{minipage}[r]{0.45\linewidth}
    \includegraphics[width=8cm,height=6cm]{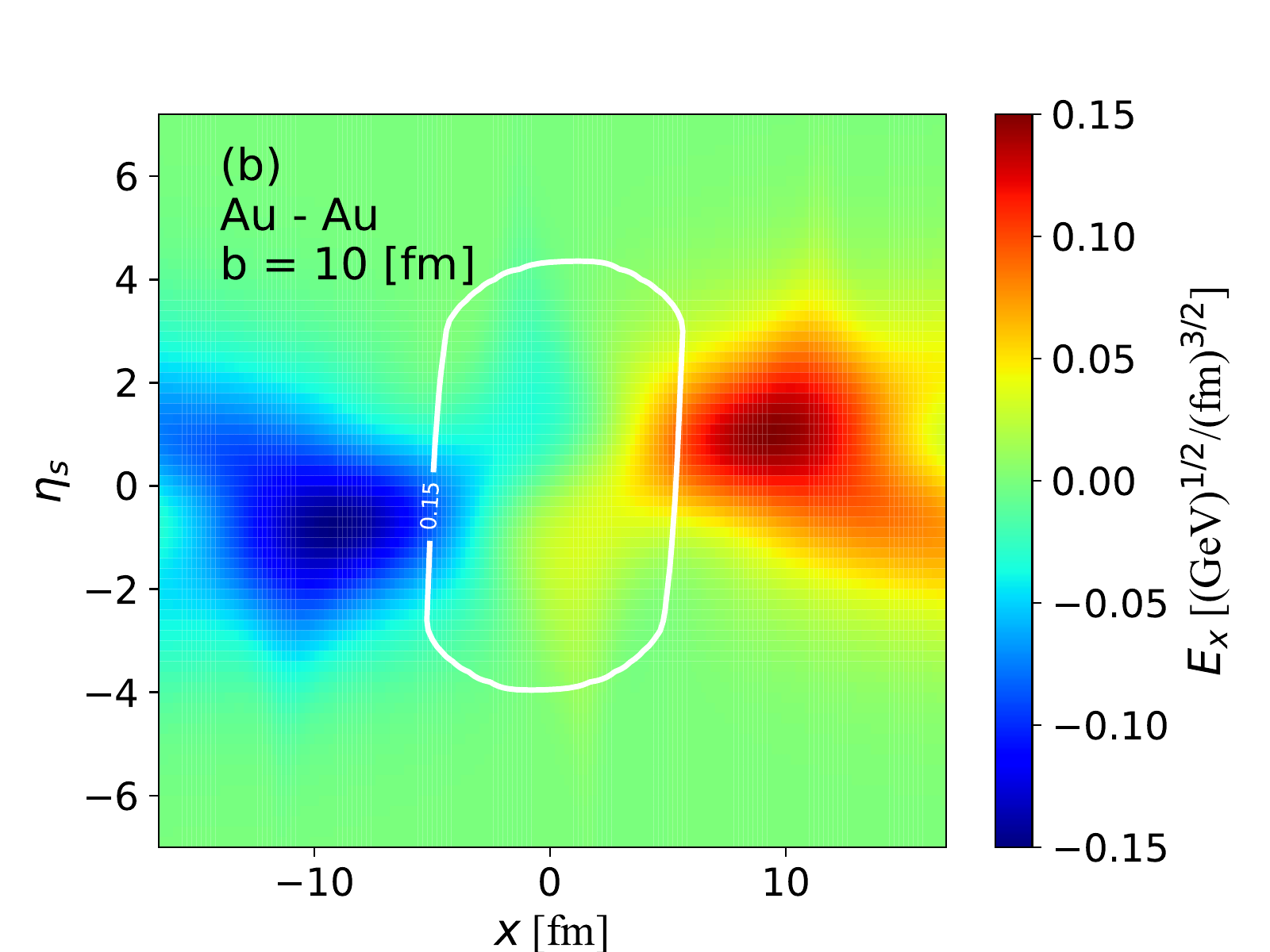}
     \end{minipage}
    \caption{(color online) The initial electromagnetic field in the reaction plane at $y = 0~\mathrm{fm}$ for Au-Au collisions. 
    We show the $y$-component of the magnetic field (a) and the $x$-component of the electric field (b), respectively.
    The white line represents the iso-thermal curve at $e = 0.15~\rm{GeV/fm^3}$.
    }
    \label{fig_ini:EM_AuAu_L}
\end{figure*}

\begin{figure*}[t]
  \begin{minipage}[l]{0.45\linewidth}
    \includegraphics[width=8cm,height=6cm]{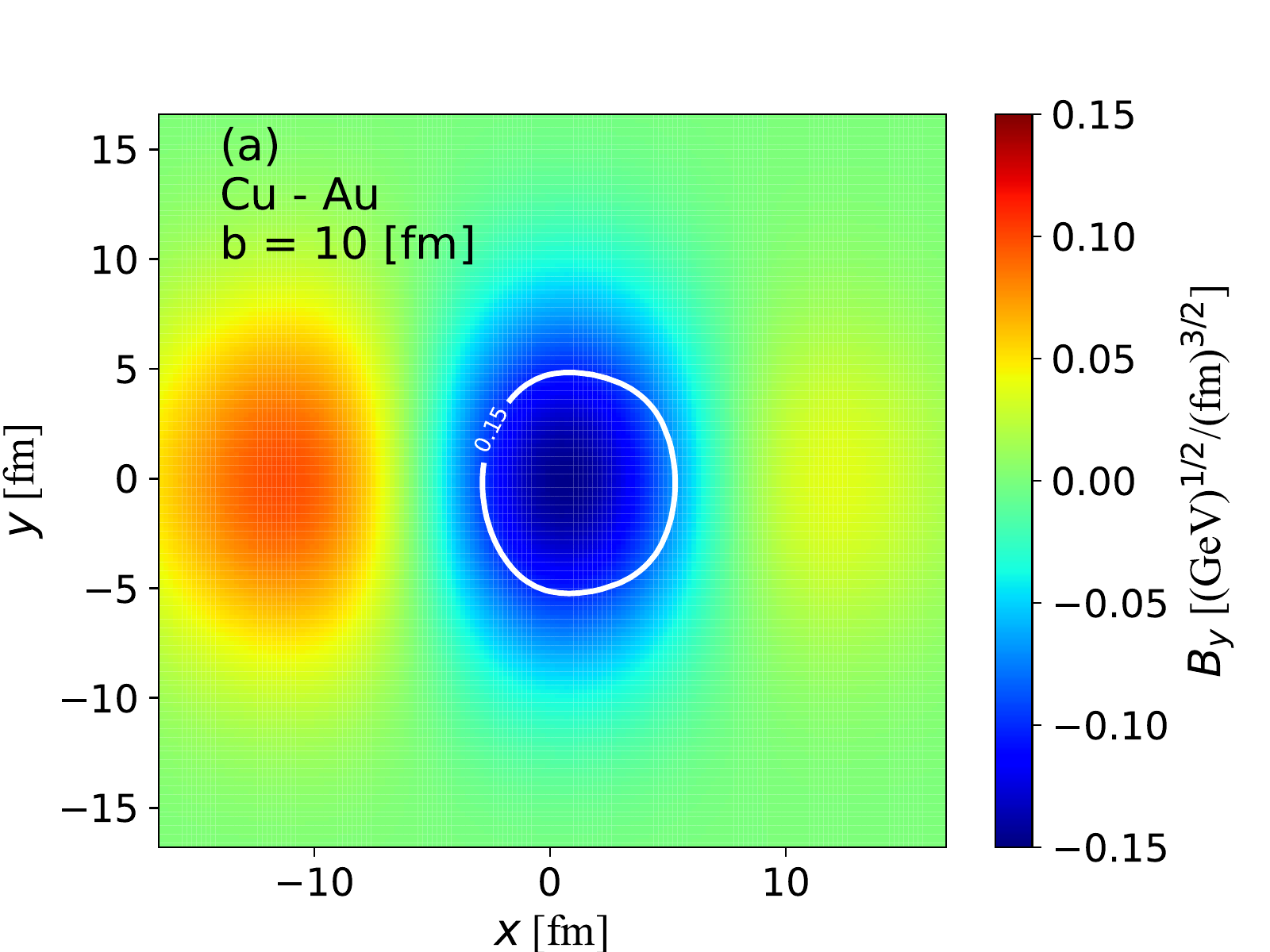}
  \end{minipage}
  \begin{minipage}[r]{0.45\linewidth}
    \includegraphics[width=8cm,height=6cm]{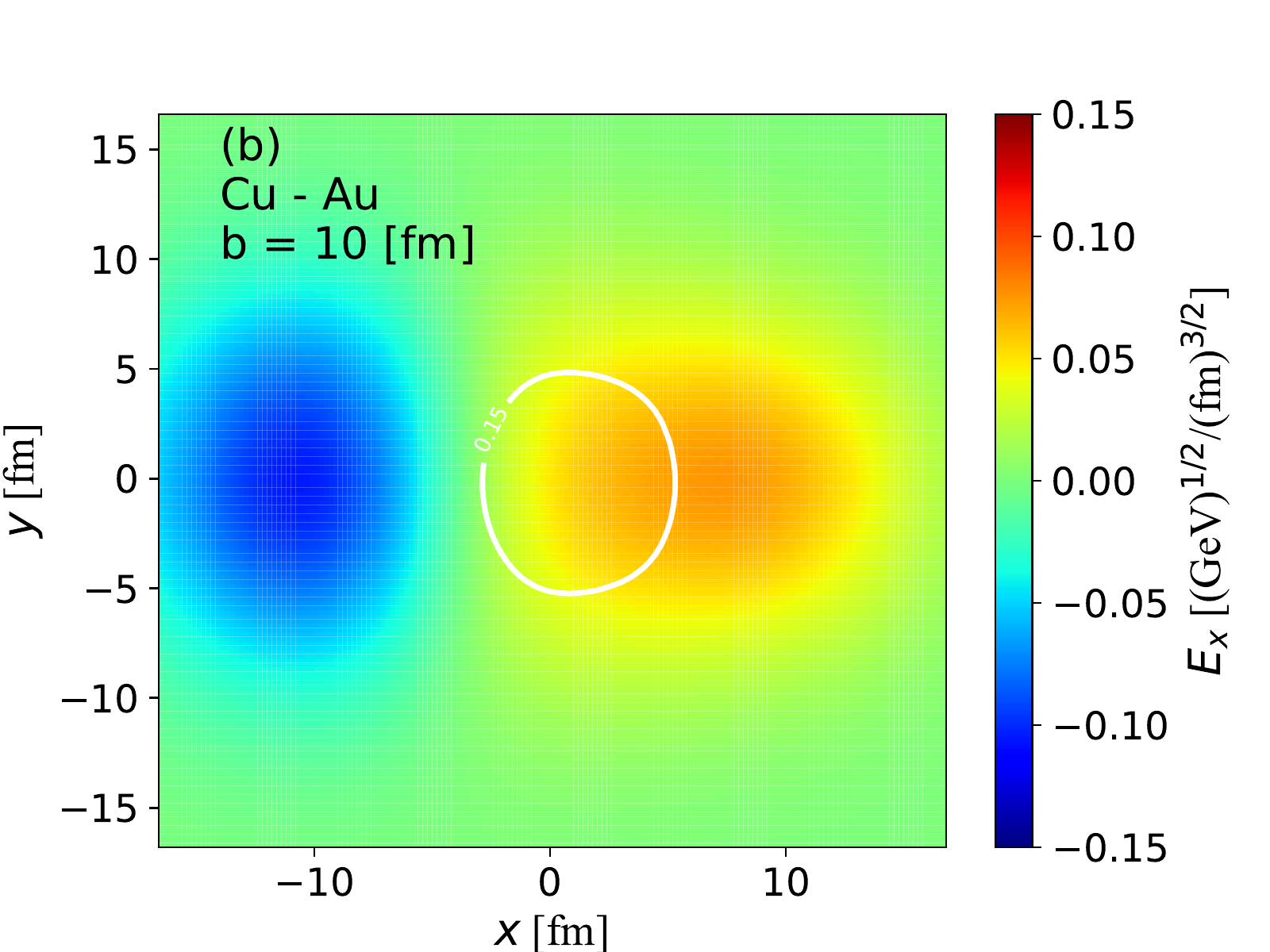}
     \end{minipage}
    \caption{(color online) The initial electromagnetic field in the transverse plane at $\eta_s = 0$ for Cu-Au collisions.
    We show the $y$-component of the magnetic field (a) and the $x$-component of the electric field (b), respectively.
    The white line represents the iso-thermal curve at $e = 0.15~\rm{GeV/fm^3}$.
    }
    \label{fig_ini:EM_CuAu_T}
\end{figure*}

\begin{figure*}[t]
  \begin{minipage}[l]{0.45\linewidth}
    \includegraphics[width=8cm,height=6cm]{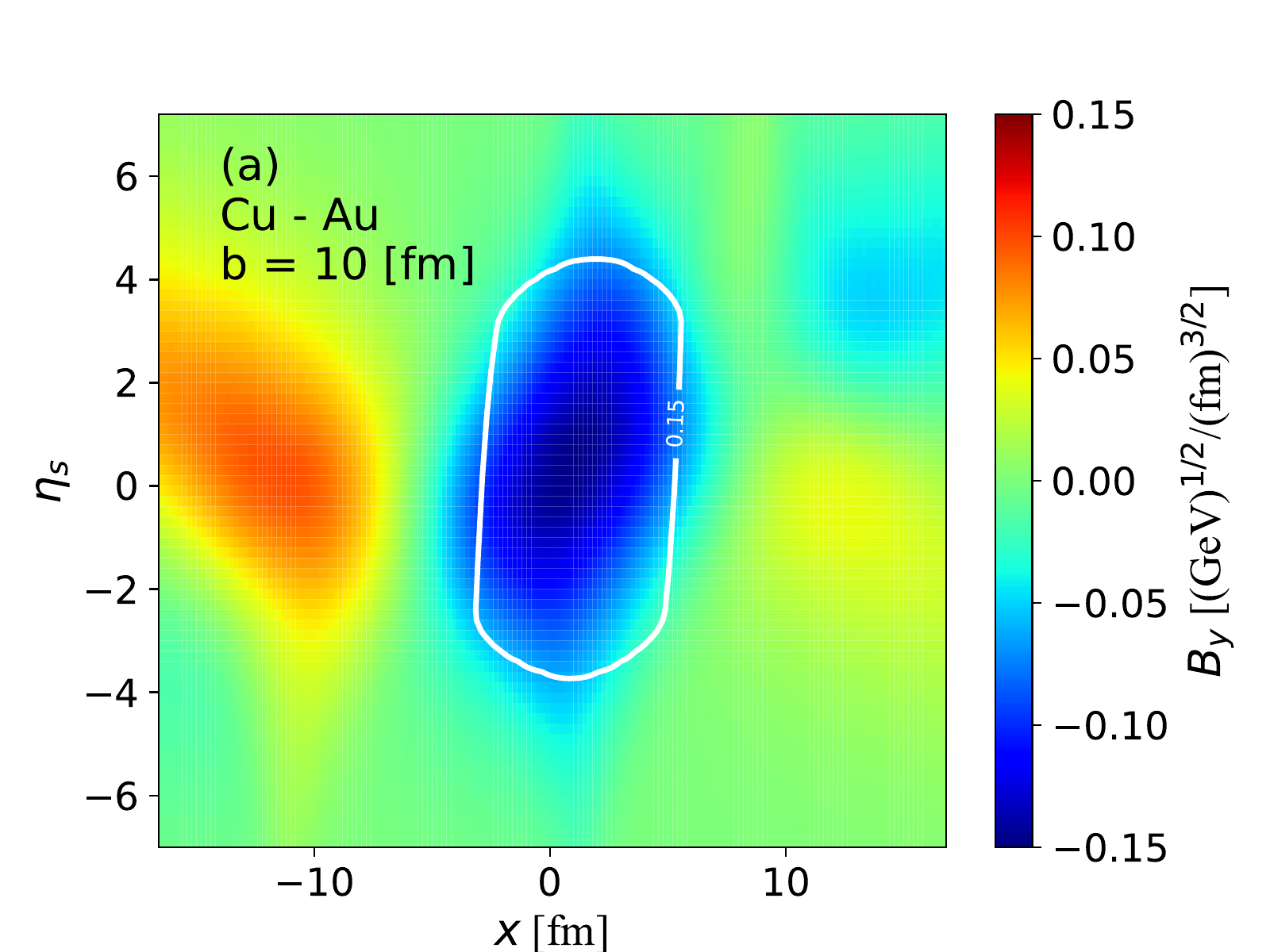}
  \end{minipage}
  \begin{minipage}[r]{0.45\linewidth}
    \includegraphics[width=8cm,height=6cm]{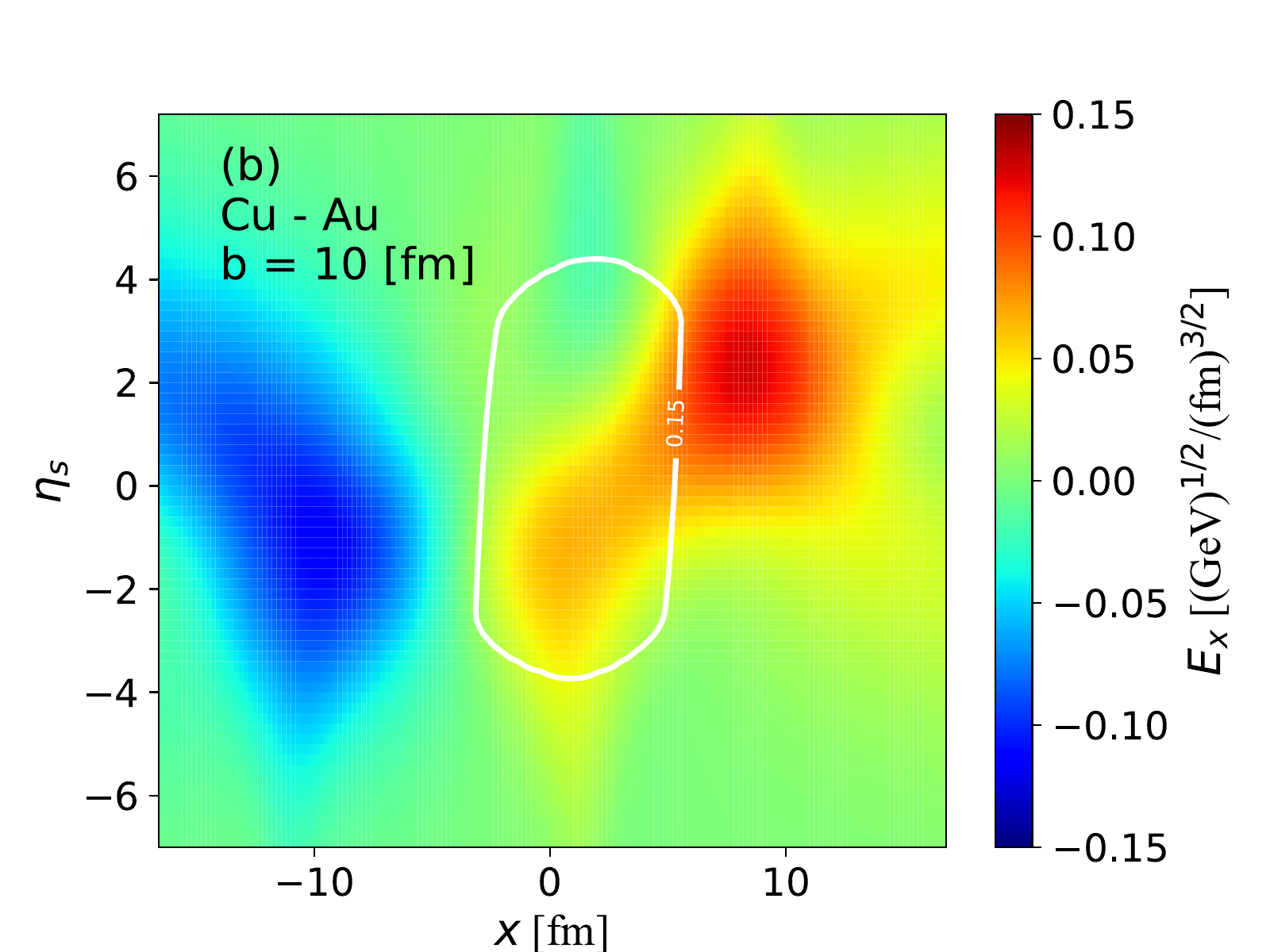}
     \end{minipage}
    \caption{(color online) The initial electromagnetic field in the reaction plane at $y=0~\mathrm{fm}$ for Cu-Au collisions.
    We display the $y$-component of the magnetic field (a) and the $x$-component of the electric field (b), respectively.
    The white line represents the iso-thermal curve at $e = 0.15~\rm{GeV/fm^3}$.
    }
    \label{fig_ini:EM_CuAu_L}
\end{figure*}

Initial conditions for the RRMHD equations are built up with the optical Glauber models~\cite{Glauber}.
The parameter selection for the initial condition of our model is based on  the ECHO-QGP simulation~\cite{Inghirami:2019mkc}.
We assume the initial energy density distribution takes the form,
\begin{equation}
   e(\mathbf{x}_\perp,\eta_s;\mathbf{b}) = e_0M(\mathbf{x}_\perp;\mathbf{b})f_{\rm{tilt}}(\eta_s),
\end{equation}
 where $e_0 = 55~\mathrm{GeV/fm}^3$~\cite{Inghirami:2019mkc} is the value of energy density at $\mathbf{x}_\perp=\mathbf{0}$ and $f_{\mathrm{tilt}}(\eta_s)$ is a longitudinal profile function with the tilted sources~\cite{PhysRevC.81.054902}. 
 The energy density distribution in the transverse plane $M(\mathbf{x}_\perp;\mathbf{b})$ is written by,
\begin{equation}
  M(\mathbf{x}_\perp;\mathbf{b}) = \frac{(1-\alpha_{\rm{H}})n_{\rm{part}}(\mathbf{x}_\perp;\mathbf{b}) + \alpha_{\rm{H}} n_{\rm{coll}}(\mathbf{x}_\perp;\mathbf{b})}{(1-\alpha_{\rm{H}})n_{\rm{part}}(\mathbf{0};\mathbf{0}) + \alpha_{\rm{H}} n_{\rm{coll}}(\mathbf{0};\mathbf{0})},
\end{equation}  
where $\mathbf{b}$ is an impact parameter, and $\alpha_{\rm{H}} = 0.05$~\cite{Inghirami:2019mkc} is the collision hardness parameter.
We have introduced a participant's number density $n_{\mathrm{part}}(\mathbf{x}_{\perp};\mathbf{b})$ and the binary nucleon collision number density $n_{\mathrm{coll}}(\mathbf{x}_{\perp};\mathbf{b})$.
We take the inelastic nucleon-nucleon cross section $\sigma_{\mathrm{NN}}^{\mathrm{inel}} = 40~\mathrm{mb}$~\cite{Inghirami:2019mkc}.
We consider the Woods-Saxon distribution as a nucleon density profile of colliding nuclei.

In the longitudinal direction, we smoothly connect the energy density distribution from the central rapidity region to the forward and backward rapidity regions by function $f_{\mathrm{tilt}}(\eta_s)$ with tilted sources of the directed flow.
We adopt the energy density distribution by the following tilted initial energy density distribution~\cite{PhysRevC.81.054902},
\begin{equation}
  M(\mathbf{x}_\perp,\eta_s;\mathbf{b}) = \frac{(1-\alpha_{\rm{H}})W_N(\mathbf{x}_\perp,\eta_s;\mathbf{b}) + \alpha_{\rm{H}}n_{\rm{coll}}(\mathbf{x}_\perp;\mathbf{b})}{(1-\alpha_{\rm{H}})W_N(\mathbf{0},0;\mathbf{0}) + \alpha_{\rm{H}}n_{\rm{coll}}(\mathbf{0};\mathbf{0})}.
\end{equation}
We have defined the wounded nucleon's weight function $W_N(\mathbf{x}_\perp,\eta_s;\mathbf{b})$ as,
\begin{eqnarray}
  W_N(\mathbf{x}_\perp,\eta_s;\mathbf{b}) = &2&(n^\mathrm{A}_{\rm{part}}(\mathbf{x}_\perp;\mathbf{b})f_-(\eta_s) \nonumber\\  &+& n^{\mathrm{B}}_{\rm{part}}(\mathbf{x}_\perp;\mathbf{b})f_+(\eta_s)),
\end{eqnarray} 
where,
\begin{equation}
  f_-(\eta_s) =
  \begin{cases}
    1 & \text{($\eta_s < -\eta_m $)} \\
    \frac{-\eta_s + \eta_m}{2\eta_m} & \text{($-\eta_m\leq \eta_s \leq \eta_m $),} \\
    0 & \text{($\eta_s > \eta_m$)}
  \end{cases}
\end{equation}
and,
\begin{equation}
  f_+(\eta_s) =
  \begin{cases}
    0 & \text{($\eta_s < -\eta_m $)} \\
    \frac{\eta_s + \eta_m}{2\eta_m} & \text{($-\eta_m\leq \eta_s \leq \eta_m $),} \\
    1 & \text{($\eta_s > \eta_m$)}
  \end{cases}
\end{equation}
where $\eta_m = 3.36$~\cite{PhysRevC.81.054902} is a parameter.
We define the tilted longitudinal profile function $f_{\rm{tilt}}(\eta_s)$ as,
\begin{equation}
  f_{\rm{tilt}}(\eta_s) = \exp\left(\frac{-(|\eta_s| - \eta_{\rm{flat}}/2)^2}{2w_\eta^2}\theta(|\eta_s| - \eta_{\rm{flat}}/2)\right),
\end{equation}
where $w_{\eta} = 4.0$ is a parameter as a width of the gauss function in $f_\mathrm{tilt}(\eta_s)$ and $\eta_{\mathrm{\rm{flat}}} = 5.9$~\cite{Inghirami:2019mkc} is a width of plateau for the rapidity distribution.

The parameters, $\alpha_{\rm{H}}, e_0, w_\eta, \eta_{\mathrm{flat}}, \tau_0$ and $\sigma_{\mathrm{NN}}^{\mathrm{inel}}$ in the initial conditions are taken from the ECHO-QGP simulations~\cite{Inghirami:2019mkc}.
We summarize the parameter set for the initial conditions of the energy density in Tab.~\ref{tab:param}.
To extract the effects of the difference of the nucleon and charge distributions between symmetric and asymmetric collision systems, we take the same value of the parameters for both of Au-Au and Cu-Au collisions except for the parameters in the Woods-Saxon distribution.

Figure~\ref{fig_ini:T} (a) shows the initial condition of the energy density in the transverse plane at $\eta_s = 0$ for Au-Au collisions at the impact parameter 10 fm.
In Figs.~\ref{fig_ini:T} (a) and (b), the white lines stand for the iso-thermal surface at $e(\eta_s, \mathbf{x}_{\mathrm{T}}) = 0.15~\mathrm{GeV/fm}^3$ which corresponds to the freezeout hypersurface at the initial time $\tau_0 = 0.4~\rm{fm}$~\cite{Inghirami:2019mkc}.
The centers of the Au are located at the points $(x,y) = (\pm 5~\mathrm{fm},0~\mathrm{fm})$.
The almond shaped hot medium is created by the collision geometry in Au-Au collisions.
The initial condition of the energy density in the transverse plane for Cu-Au collisions is shown in Fig.~\ref{fig_ini:T} (b).
The centers of Au and Cu are located at $(x,y) = (-5~\mathrm{fm},0~\mathrm{fm})$ and $(5~\mathrm{fm},0~\mathrm{fm})$, respectively.
The effect of asymmetric collision system appears in deformation of the freezeout hypersurface.
Figures~\ref{fig_ini:L}~(a) and (b) represent the profiles of the initial energy density in the reaction plane at $y = 0~\mathrm{fm}$ for Au-Au and Cu-Au collisions, respectively.
In Cu-Au collisions, the forward rapidity corresponds to the Cu-going direction.
In the both of Au-Au and Cu-Au collisions, the tilted pressure gradient is the source of the directed flow~\cite{PhysRevC.81.054902}.
\begin{table}
 \begin{center}
 \caption{The values of parameters in the initial conditions for both of $\sqrt{s_{\mathrm{NN}}} = 200~\mathrm{GeV}$ Au-Au and Cu-Au collisions. }
 \label{tab:param}
  {\tabcolsep = 0.1cm
  {\renewcommand\arraystretch{1.8}
 \begin{tabular}{@{\extracolsep{\fill}}ll|r@{\extracolsep{\fill}}}\hline\hline
 Parameter&Description&Value\\ \hline
 $\alpha_{\rm{H}}$&Collision hardness&0.05 \\
 $e_0$&Energy density at $(\eta_s,\mathbf{x}_\mathrm{T}) =(0,\mathbf{0})$&55 [GeV/fm$^3$] \\
 $\eta_m$&Slope of the tilted source & 3.36\\
 $\eta_{\mathrm{flat}}$&Width of the plateau&5.9\\
 $w_\eta$&Width of the gauss function&0.4\\
 $\tau_0$&Initial time&0.4 [fm]\\
 $\sigma^{\mathrm{inel}}_{\mathrm{NN}}$&Inelastic cross section& 40 [mb]\\\hline\hline
\end{tabular}
  }
  }
\end{center}
\end{table}

\subsection{The initial electromagnetic field}\label{2}
We compute initial electromagnetic fields based on Ref.~\cite{PhysRevC.88.024911}.
We consider the electromagnetic fields produced by the electric charge $e$ moving along parallel to the beam axis ($\hat{\mathbf{z}}$) with velocity $v$ in the laboratory frame by an observer located at $\mathbf{r} = z\hat{\mathbf{z}} + \mathbf{x}_\perp$ in the Minkowski coordinates.
Such a system follows the Maxwell equations with the source term of point charged particles moving in the direction of the beam axis ($\hat{\mathbf{z}}$),
\begin{gather}
 \nabla\cdot \mathbf{B} = 0,~~~\nabla\times \mathbf{E} = -\frac{\partial \mathbf{B}}{\partial t},\\
\nabla\cdot \mathbf{D} = e\delta(z-vt)\delta(\mathbf{b}),\\
\nabla\times \mathbf{H} = \frac{\partial \mathbf{D}}{\partial t} + \sigma_0 \mathbf{E} + ev\hat{\mathbf{z}}\delta(z-vt)\delta(\mathbf{b}),
\end{gather}
where $\mathbf{H} = \mu\mathbf{B}$ and $\mathbf{D} = \epsilon\mathbf{E}$.
In the case of $\gamma_0\sigma_0 \rm{b} \gg 1$, Maxwell equations reduce simple solutions by integration,
\begin{eqnarray}
  E_r = B_\phi = \frac{e(\hbar c)^{3/2}}{2\pi}\frac{\mathrm{b}\sigma_0/(\hbar c)}{4x_\pm^2}\exp\left(-\frac{\mathrm{b}^2\sigma_0/(\hbar c)}{4x_\pm}\right)\nonumber,\\
  E_z = -\frac{e(\hbar c)^{3/2}}{4\pi}\frac{x_\pm - \mathrm{b}^2\sigma_0/(4\hbar c)}{\gamma_0^2x_\pm^3}\exp\left(-\frac{\mathrm{b}^2\sigma_0/(\hbar c )}{4x_\pm}\right),\nonumber\\
\end{eqnarray}
where we define $\gamma_0 = 1/\sqrt{1 - v^2}$ and $x_{\pm} = t\pm v/z$.
We assume a constant permittivity $\epsilon = 1$, a constant permeability $\mu = 1$ and a constant finite electrical conductivity $\sigma_0 = 5.8~\mathrm{MeV}$~\cite{Ding:2010ga,Aarts:2007wj}.
To clarify the dimension of electromagnetic fields, $\mathrm{GeV^{1/2}/fm^{3/2}}$, we explicitly write $\hbar$ and $c$.
We take the electric charge distribution inside two colliding nuclei as being uniform and spherical for simplicity.
Then total electromagnetic fields are derived by integration over the interior of colliding nuclei in each point of our computational grid.

We show the profile of electromagnetic fields in the transverse and reaction planes for Au-Au collisions in Figs.~\ref{fig_ini:EM_AuAu_T} and \ref{fig_ini:EM_AuAu_L}, respectively.
In Fig.~\ref{fig_ini:EM_AuAu_T} (a), the $y$-component of the magnetic field inside the freezeout hypersurface (white line) is stronger than that outside the freezeout hypersurface by the Biot-Savart law.
In Fig.~\ref{fig_ini:EM_AuAu_T} (b), the $x$-component of the electric field created by the two nuclei cancels each other and becomes zero around $(x,y,\eta_s) = (0~\mathrm{fm},0~\mathrm{fm},0)$ by the symmetric charge distribution inside colliding nuclei.
Figure~\ref{fig_ini:EM_AuAu_L}~(a) shows the $y$-component of the magnetic field in the reaction plane. 
We can see that, inside the medium, the $y$-component of the magnetic field is finite. 
Figure~\ref{fig_ini:EM_AuAu_L}~(b) represents the $x$-component of the electric field in the reaction plane.
If we focus on the behavior of the $x$-component of the electric field as a function of $\eta_s$ around $x\sim 0$ fm, it has a positive value in the backward rapidity, decreases with $\eta_s$, it becomes vanishing at $\eta_s = 0$, and has a negative value in the forward rapidity. 
This indicates that the electric field produced colliding nuclei is canceled out each other at $\eta_s = 0$.

The profiles of electromagnetic fields for Cu-Au collisions are shown in Figs.~\ref{fig_ini:EM_CuAu_T} and~\ref{fig_ini:EM_CuAu_L}.
In Fig.~\ref{fig_ini:EM_CuAu_T}~(a), the distribution of the $y$-component of the magnetic field is similar to that in Au-Au collisions.
However, because of a difference between the charge density of Cu and that of Au, the magnetic field in the $x > 5$ fm region is smaller than that in $x < -5$ fm.
In Fig.~\ref{fig_ini:EM_CuAu_T} (b), we observe the asymmetric profile of the $x$-component of the electric field which is different from symmetric profile in Au-Au collisions in Fig.~\ref{fig_ini:EM_AuAu_T}~(b).
The non-zero $x$-component of the electric field exists inside the freezeout hypersurface.
The magnitude of the electric field on the Cu side ($x > 0~\mathrm{fm}$) is larger than that on the Au side ($x < 0~\mathrm{fm}$).
Figures~\ref{fig_ini:EM_CuAu_L} (a) and (b) show the $y$-component of the magnetic field and the $x$-component of the electric field in the reaction plane, respectively.
The $y$-component of the magnetic field on the Au side ($x > 7~\mathrm{fm}$) is larger than that on the Cu side ($x < -7~\mathrm{fm}$) as shown in Fig.~\ref{fig_ini:EM_CuAu_T}~(a). 
In Fig.~\ref{fig_ini:EM_CuAu_L}~(b), the initial electric field in the Au-going ($\eta_s < 0$) side is larger than that in the Cu-going ($\eta_s > 0$) side because the electric field created by the Au is dominated.
The characteristic features of the initial electromagnetic field in Cu-Au collisions may affect collective flows, in particular, asymmetric flows.

\section{Numerical results}\label{results}
\subsection{Relativistic resistive magneto-hydrodynamic expansion}

\begin{figure*}[t]
  \begin{minipage}[l]{0.45\linewidth}
    \includegraphics[width=9cm,height=7cm]{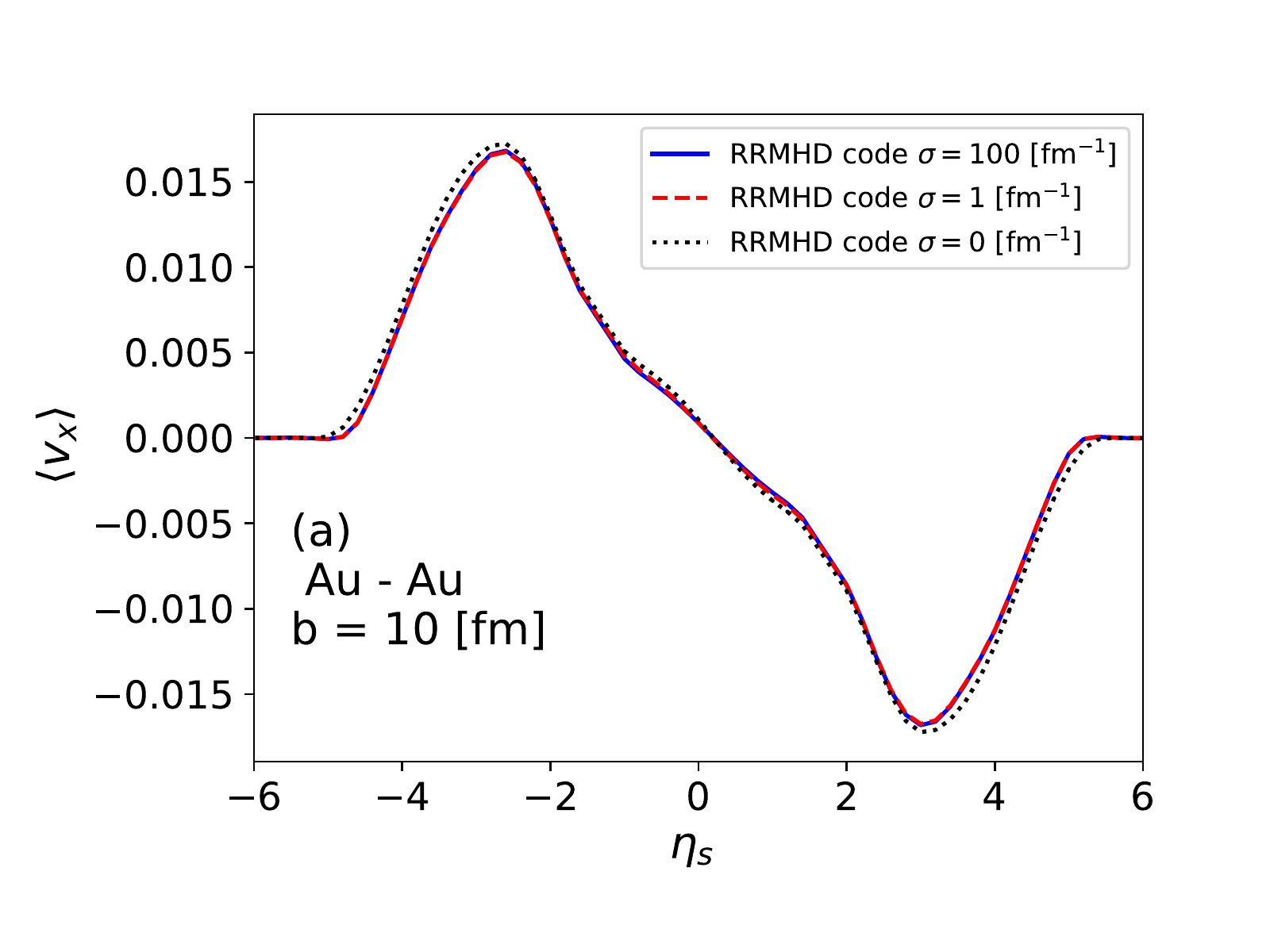}
  \end{minipage}
  \begin{minipage}[r]{0.45\linewidth}
    \includegraphics[width=9cm,height=7cm]{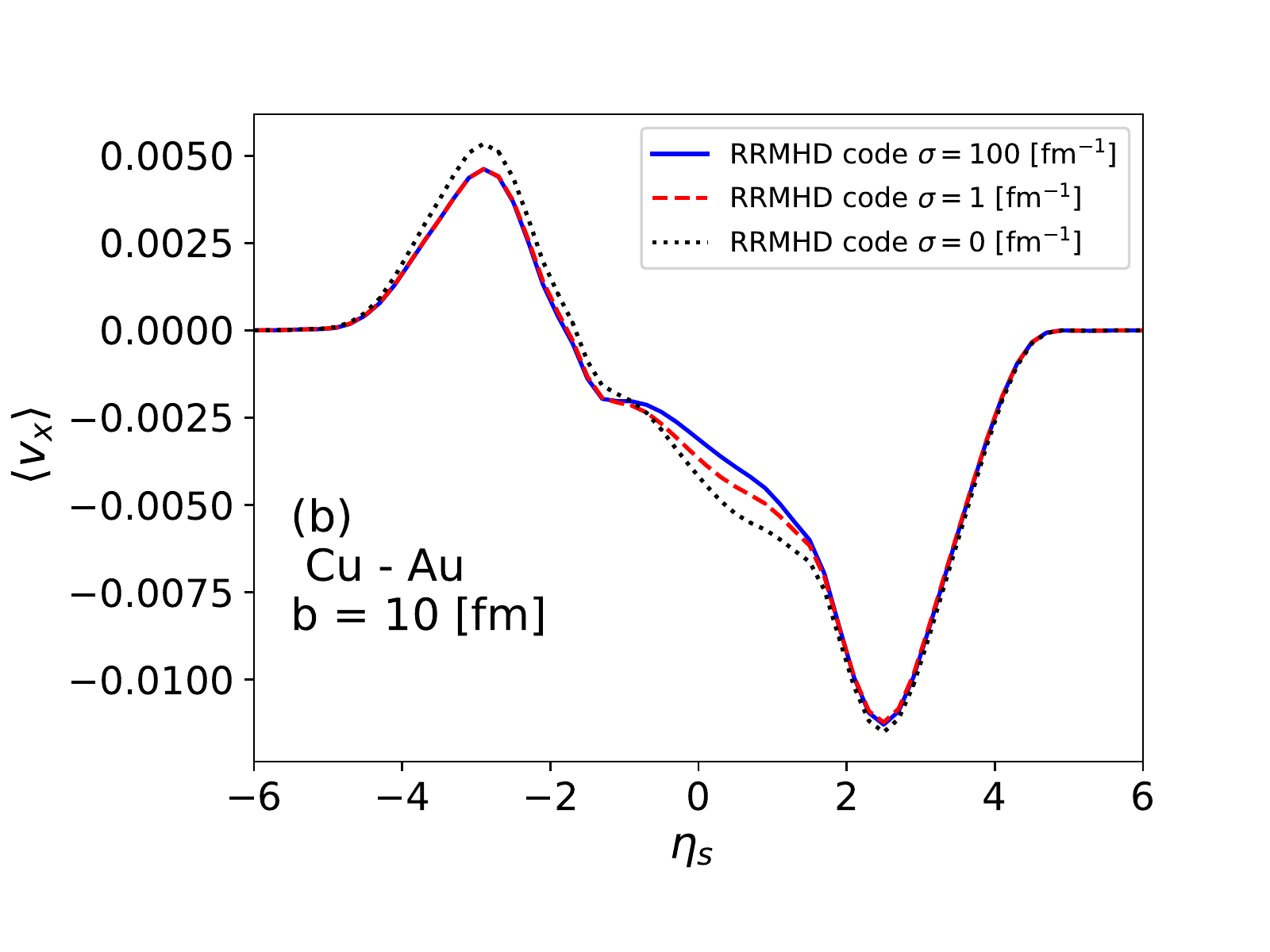}
     \end{minipage}
    \caption{(color online) The space averaged flow in the $x$-direction as a function of space rapidity at $\tau = 3.0~\mathrm{fm}$. 
    The blue solid, red dashed, black dotted lines show $\sigma = 100, 1$ and $0~\rm{fm^{-1}}$, respectively.
    We show the cases of Au-Au collisions (a) and Cu-Au collisions (b) at $\sqrt{s_{\rm{NN}}} = 200~\rm{GeV}$.
    }
    \label{fig_hydro:vx}
\end{figure*}

The RRMHD simulation is performed for the tilted initial conditions with electromagnetic fields in Au-Au and Cu-Au collisions at $\sqrt{s_{\rm{NN}}}=200~\rm{GeV}$.
We start the RRMHD simulation at initial time $\tau_0 = 0.4~\mathrm{fm}$.
Figures~\ref{fig_hydro:vx}~(a) and (b) show the velocity profile ($\langle v_x \rangle$) as a function of $\eta_s$ in the cases of $\sigma = 0,1$ and $100~\mathrm{fm}^{-1}$ at time $\tau = 3.0$ fm.
The definition of $\langle v_x \rangle$ is given by,
\begin{equation}
  \langle v_x \rangle = \frac{\int dydx \gamma e(x,y,\eta_s)v_x(x,y,\eta_s)}{\int dydx \gamma e(x,y,\eta_s)}.
\end{equation} 
The blue solid, red dashed and black dotted lines stand for $\langle v_x \rangle$ for $\sigma =  100, 1,$ and $0$ $\rm{fm}^{-1}$, respectively.
Our simulation with zero electrical conductivity is equivalent to the relativistic ideal hydrodynamics simulation. 

The Au-Au collision system case is shown in Fig.~\ref{fig_hydro:vx} (a).
The $\eta_s$ dependence of velocity appeared in $|\eta_s| < 3$ is reflected from the tilted sources in the initial conditions.
There are only small differences of the profile of the velocity among the electrical conductivities.
However, if we focus on the rapidity region $|\eta_s| < 2 $, the fluid velocity is slightly suppressed in the forward and backward rapidity regions by existence of the electromagnetic fields with finite electrical conductivity.
For example, $\langle v_x \rangle$ with $\sigma = 100~\mathrm{fm}^{-1}$ is less than that with $\sigma = 0~\mathrm{fm}^{-1}$.
The difference of $\langle v_x \rangle$ is evaluated by $|\Delta \langle v_x \rangle| := |\langle v_x \rangle_{\sigma = 100~\mathrm{fm}^{-1}} -\langle v_x \rangle_{\sigma= 0~\mathrm{fm}^{-1}}| \sim 0.7 \times 10^{-3}c$ at $\eta_s = -1.0$.
 
As shown in Fig.~\ref{fig_ini:EM_AuAu_L}~(b), the $x$-component of the electric field is finite in the forward and backward rapidity regions inside the freezeout hypersurface.
The energy of this electric field is converted to the fluid energy by the dissipation associated with Ohmic conduction $\sigma \mathbf{E}\cdot\mathbf{E}$.
After just one time step of RRMHD simulation ($\Delta\tau = 0.02~\mathrm{fm}$) from the initial condition in Fig.~\ref{fig_ini:L}~(a), this dissipation makes the pressure gradient of the medium around $(\eta_s,x)=(-1.0,2.5~\mathrm{fm})$ with $\sigma = 100~\mathrm{fm^{-1}}$ flatter ($\sim 0.8 \times 10^{-3}~\mathrm{GeV/fm}^{4}$) than that with $\sigma = 0~\mathrm{fm^{-1}}$.
If we ignore the Maxwell's stress tensor, the RRMHD equations part in Eq.~(\ref{conservative form}) contain the equation $\partial_\tau u_x = -\frac{1}{e + p}\partial_x p$.
This means that $\Delta\langle v_x \rangle$ is proportional to the difference of the pressure gradient.
Therefore, the reduction of the fluid velocity is the same order of the difference of the pressure gradient between $\sigma = 100$ and $0~\mathrm{fm}^{-1}$ cases.
We note that the contribution of the Maxwell's stress force is very small because of the large value of plasma beta ($\beta\sim 1000$) in the freezeout hypersurface.
Here, the plasma beta is defined by the ratio of bulk pressure to magnetic pressure $\beta = p/p_{\mathrm{em}}$.

In the zero electrical conductivity case, the fluid is completely decoupled with electromagnetic fields.
The profile of velocity with zero electrical conductivity (black dotted line) is consistent with the result of the relativistic ideal hydrodynamic simulation in the initial condition with the tilted sources which corresponds to Fig.~6 in Ref.~\cite{PhysRevC.81.054902}.

The Cu-Au collision system case is shown in Fig.~\ref{fig_hydro:vx} (b).
The electrical conductivity dependence of $\langle v_x \rangle$ is clearly observed around $\eta_s = 0$; the amplitude of $\langle v_x \rangle$ decreases with electrical 
conductivity.
In particular, at $\eta_s = 0$, $\langle v_x \rangle$ with $\sigma = 100~\mathrm{fm}^{-1}$ is less than that with $\sigma = 0~\mathrm{fm}^{-1}$.
The difference between $\langle v_x\rangle$ with $\sigma = 100$ $\mathrm{fm}^{-1}$ and that with $\sigma = 0$ $\mathrm{fm}^{-1}$ is $|\Delta\langle v_x \rangle| \sim 0.0025c$.
As shown in the initial condition of the electric field in Fig.~\ref{fig_ini:EM_CuAu_T}~(b), the electric current is induced in the $x$-direction.
The Ohm's law converts the energy from the electric field to the fluid.
After just one time step from the initial condition, the pressure gradient of QGP medium around $(\eta_s,x) = (0,1.5~\mathrm{fm})$ with $\sigma = 100~\mathrm{fm^{-1}}$ becomes flatter ($\sim 0.003~\mathrm{GeV/fm}^{4}$) than that with $\sigma = 0~\mathrm{fm^{-1}}$.
The difference of the pressure gradient is the same order of the reduction of the fluid flow. 
Furthermore, the contribution of the Maxwell's stress force is not visible since the plasma beta is also large in the asymmetric collision system.
The electrical conductivity dependence is the result of the energy transfer from electromagnetic fields to the fluid by the dissipation.
This reduction is larger than that of Au-Au collisions.
This reason is that, inside the freezeout hypersurface in Fig.~\ref{fig_ini:EM_CuAu_T}, the electric field in Cu-Au collisions has larger value than that in Au-Au collisions.
It indicates that the large Ohmic conduction is induced in asymmetric collisions.

\begin{figure*}[t]
  \begin{minipage}[l]{0.45\linewidth}
    \includegraphics[width=9cm,height=7cm]{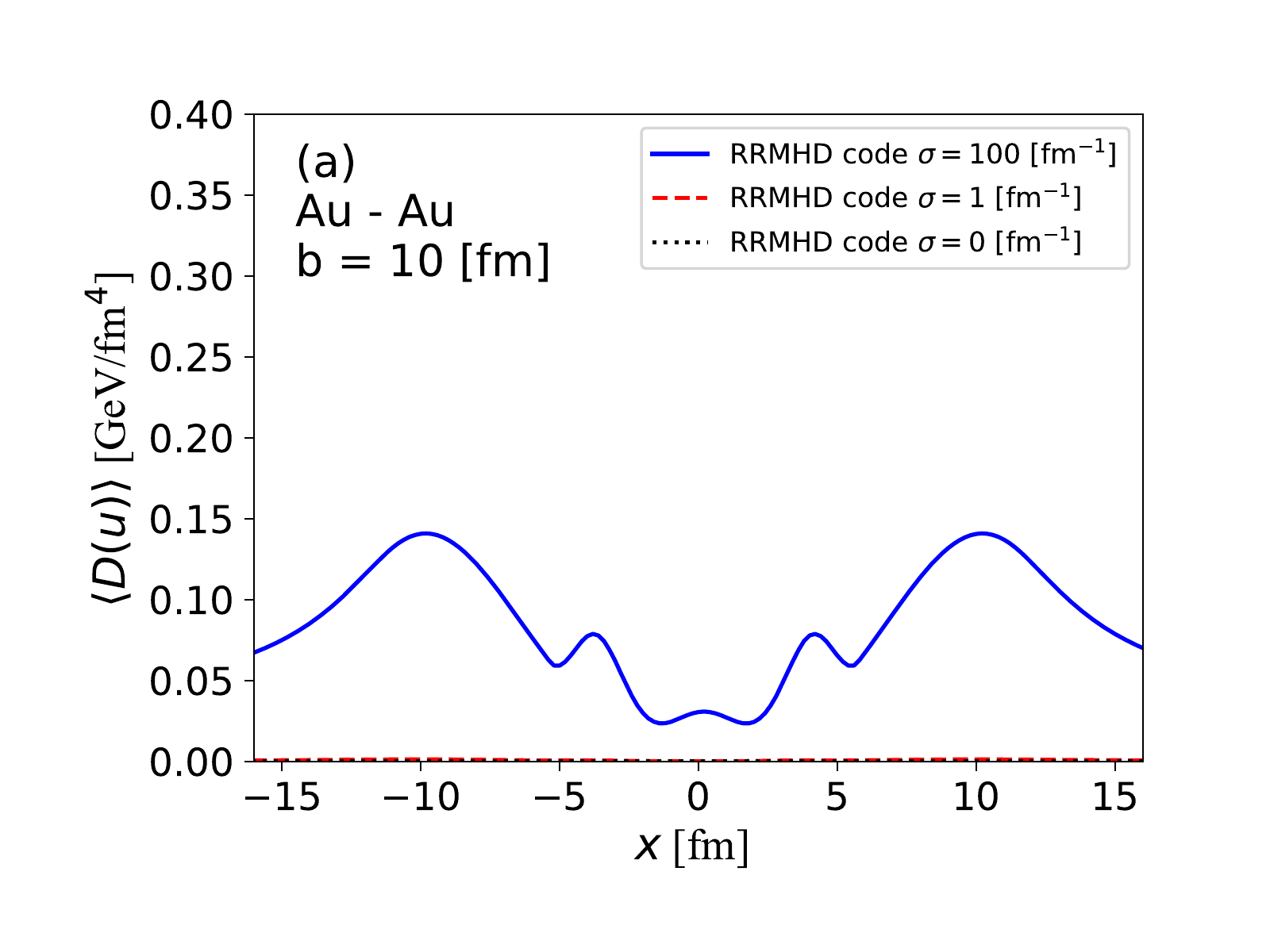}
  \end{minipage}
  \begin{minipage}[r]{0.45\linewidth}
    \includegraphics[width=9cm,height=7cm]{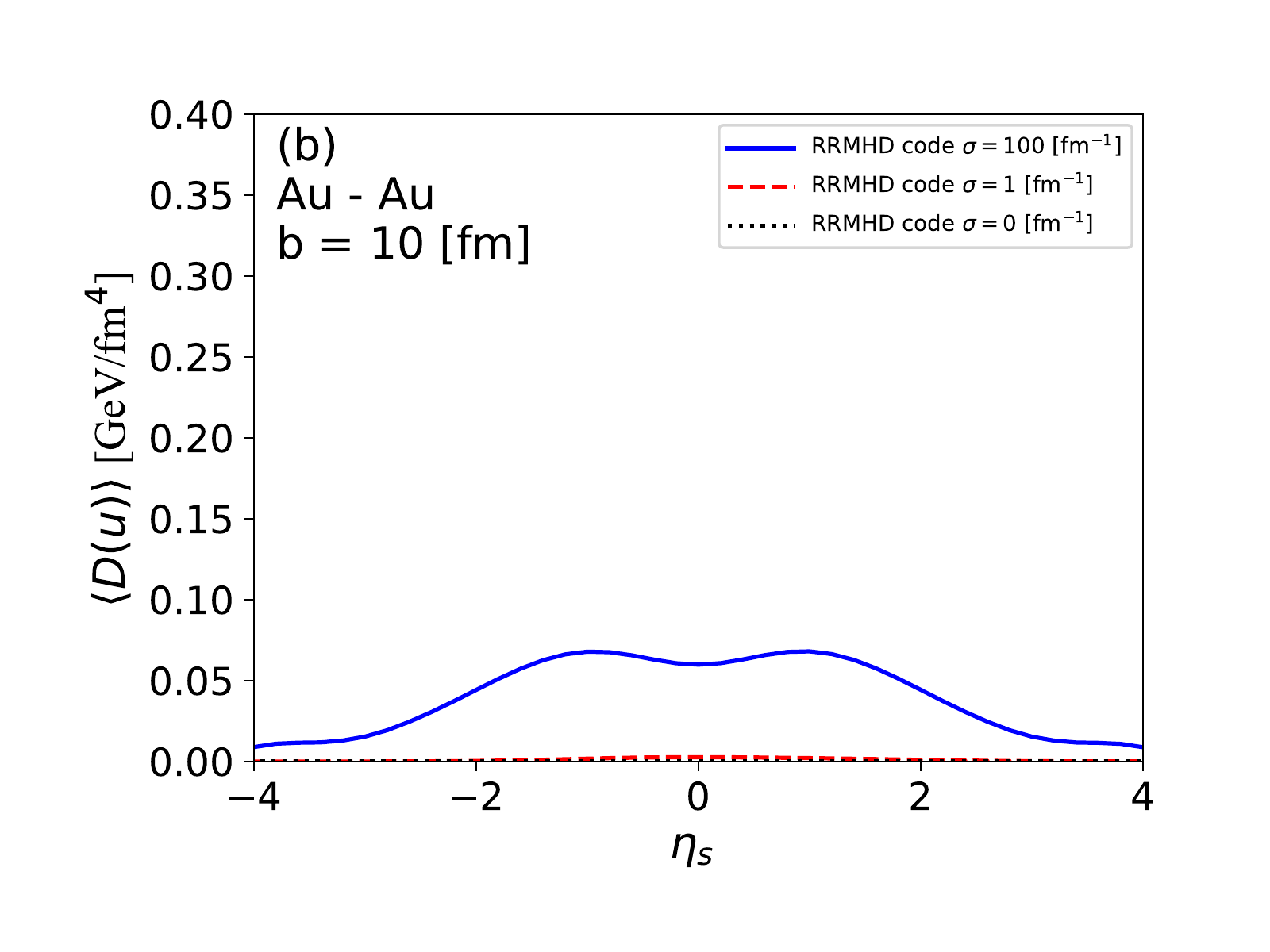}
     \end{minipage}
    \caption{(color online) The weighted dissipation measure as a function of $x$ (a) and as a function of $\eta_s$ (b) for Au-Au collisions at initial time ($\tau = 0.4$ $\rm{fm}$).
    The blue solid, red dashed, black dotted lines show $\sigma = 100, 1$ and $0~\rm{fm^{-1}}$, respectively.
    }
    \label{fig_hydro:Du_AuAu}
\end{figure*}

\begin{figure*}[t]
  \begin{minipage}[l]{0.45\linewidth}
    \includegraphics[width=9cm,height=7cm]{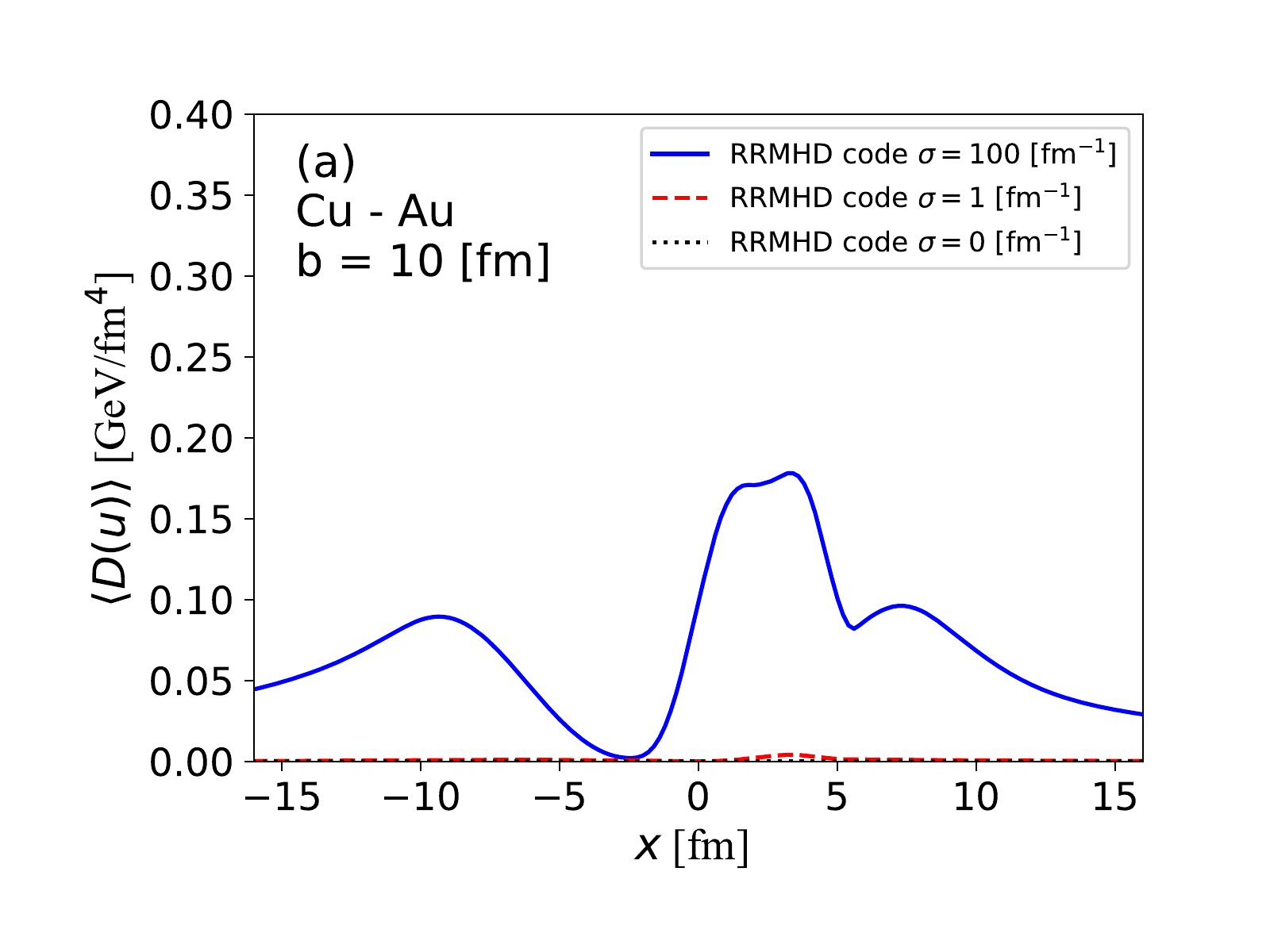}
  \end{minipage}
  \begin{minipage}[r]{0.45\linewidth}
    \includegraphics[width=9cm,height=7cm]{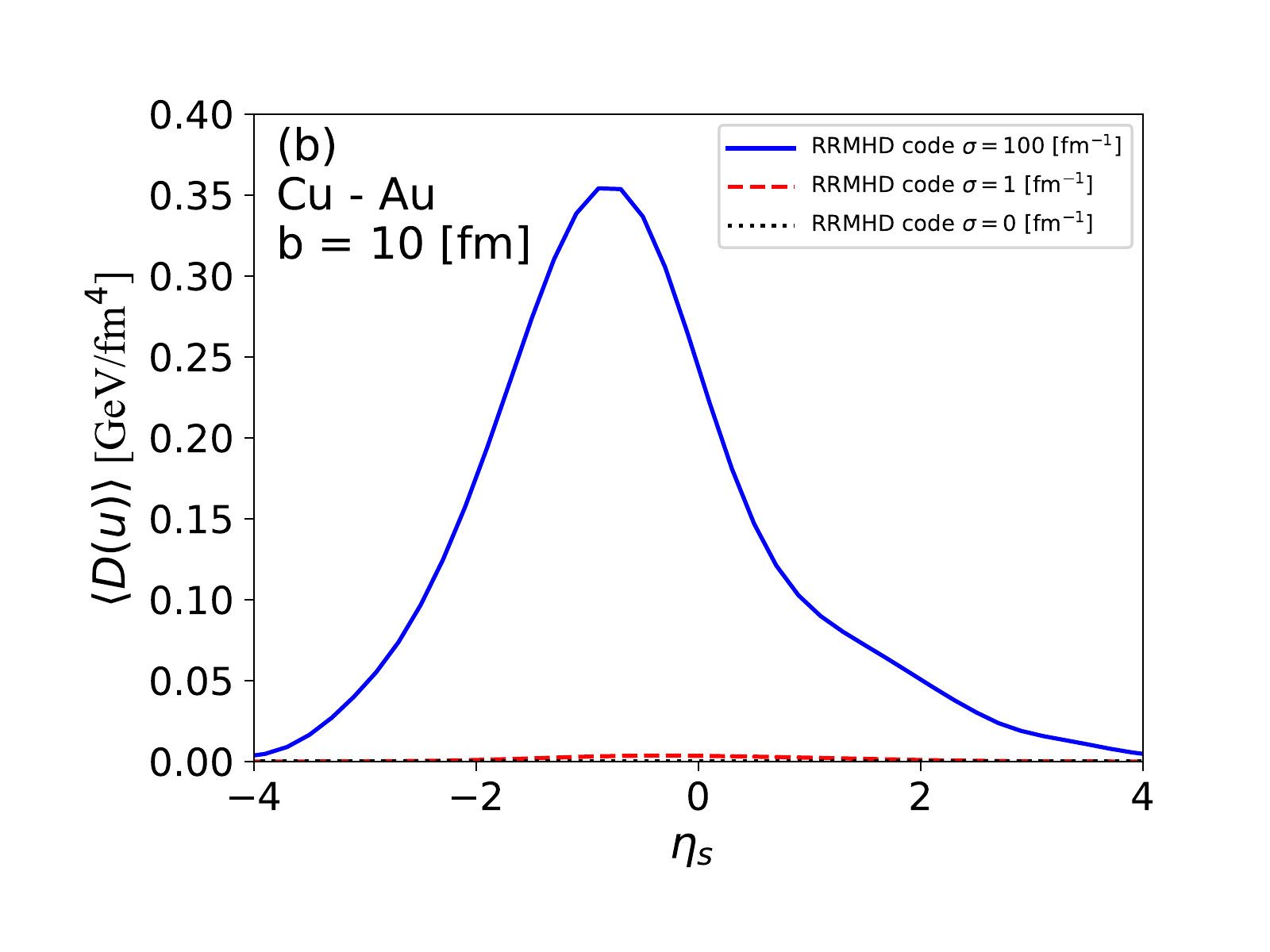}
     \end{minipage}
    \caption{(color online) The weighted dissipation measure as a function of $x$ (a) and as a function of $\eta_s$ (b) for Cu-Au collisions at the initial time ($\tau = 0.4$ $\rm{fm}$).
    The blue solid, red dashed, black dotted lines show $\sigma = 100, 1$ and $0~\rm{fm^{-1}}$, respectively.
    }
    \label{fig_hydro:Du_CuAu}
\end{figure*}

To make it clear that the energy transfer from electromagnetic fields to the fluid occurs, we introduce the dissipation measure $D(u)~\rm{GeV}/\rm{fm}^4$~\cite{PhysRevLett.106.195003} defined as,
\begin{eqnarray}
  D(u) &=& j^\mu e_\mu \nonumber\\
       &=& \gamma[\mathbf{j}\cdot(\mathbf{E}+\mathbf{v}\times\mathbf{B})-q(\mathbf{v}\cdot\mathbf{E})].
\end{eqnarray} 
The dissipation measure was first introduced in Ref.~\cite{PhysRevLett.106.195003} to detect the dissipation region in collisionless magnetic reconnection. 
This quantity represents the conversion rate of the energy from the electromagnetic field to the fluid by the dissipation. 

Figure~\ref{fig_hydro:Du_AuAu}~(a) shows the weighted spatial distributions of the dissipation measure $D(u)$ as a function of $x$ in Au-Au collisions at initial time $\tau_0 = 0.4~\mathrm{fm}$,
\begin{equation}
  \langle D(u) \rangle(x) = \frac{\int dyd\eta_s \gamma e(x,y,\eta_s)D(u)(x,y,\eta_s)}{\int dyd\eta_s \gamma e(x,y,\eta_s)}.
\end{equation}
The blue solid, red dashed and black dotted lines stand for $\langle D(u) \rangle$ for $\sigma =  100, 1,$ and $0$ $\rm{fm}^{-1}$, respectively.
Since the dissipation measure is proportional to the electrical conductivity, the amplitude of $\langle D(u) \rangle $ with $\sigma = 1$ $\mathrm{fm}^{-1}$ is $10^{-2}$ times smaller than that with $\sigma = 100~\mathrm{fm^{-1}}$.
The timescale of the energy transfer by dissipation is determined by the electrical conductivity, $\tau_\sigma\sim 1/\sigma$.
In other words, the energy transfer instantaneously occurs in the high conductive case, whereas it gradually occurs in the resistive case.
The $\langle D(u)\rangle$ has a symmetric structure for the positive $x$ and the negative $x$.
It becomes small in the region of $|x| < 3$ fm, which is reflected from the small initial electric fields around $x\sim 0$ in Fig.~\ref{fig_ini:EM_AuAu_T} (b).
Near the freezeout hypersurface around $|x|\sim 5$ fm, there are two peaks.
Outside the medium, $|x|\sim 10$ fm, two large peaks exist, however they do not give influence to the time evolution of the fluid.
The symmetric structure of $\langle D(u) \rangle$ suggests that the converted electromagnetic energy may not affect the directed flow, but may change the amplitude of elliptic flow.
In Fig.~\ref{fig_hydro:Du_AuAu}~(b), the weighted spatial distributions of the dissipation measure $D(u)$ as a function of $\eta_s$,  
\begin{equation}
\langle D(u) \rangle(\eta_s) = \frac{\int dydx \gamma e(x,y,\eta_s)D(u)(x,y,\eta_s)}{\int dydx \gamma e(x,y,\eta_s)},
\end{equation}
are shown.
The profile of $\langle D(u)\rangle$ is understood by integration of the initial electric fields over $x$ in Fig.~\ref{fig_ini:EM_AuAu_L} (b).

Figure~\ref{fig_hydro:Du_CuAu}~(a) represents the weighted dissipation measure as a function of $x$ in the case of Cu-Au collisions.
In contrast to the symmetric collision, the symmetric structure for the positive $x$ and the negative $x$ is broken.
There are the QGP medium in the region of $-2 < x < 5$ fm where $\langle D(u) \rangle$ has only one peak around $x\sim 3$ fm in side of Cu.
In addition, the magnitude of the peak of $\langle D(u) \rangle$ is larger than the two peaks in the symmetric collision.
The profile of the dissipation measure suggests that the energy transfer by Ohm's law alters behavior of the directed flow.
The $\eta_s$ dependence of the weighted dissipation measure has the asymmetric profile as represented in Fig.~\ref{fig_hydro:Du_CuAu}~(b).
The largest peak of the dissipation measure is located at $\eta_s \sim -1$.

\subsection{The directed flow}
\begin{figure*}[t]
  \begin{minipage}[l]{0.48\linewidth}
    \includegraphics[width=9cm,height=7cm]{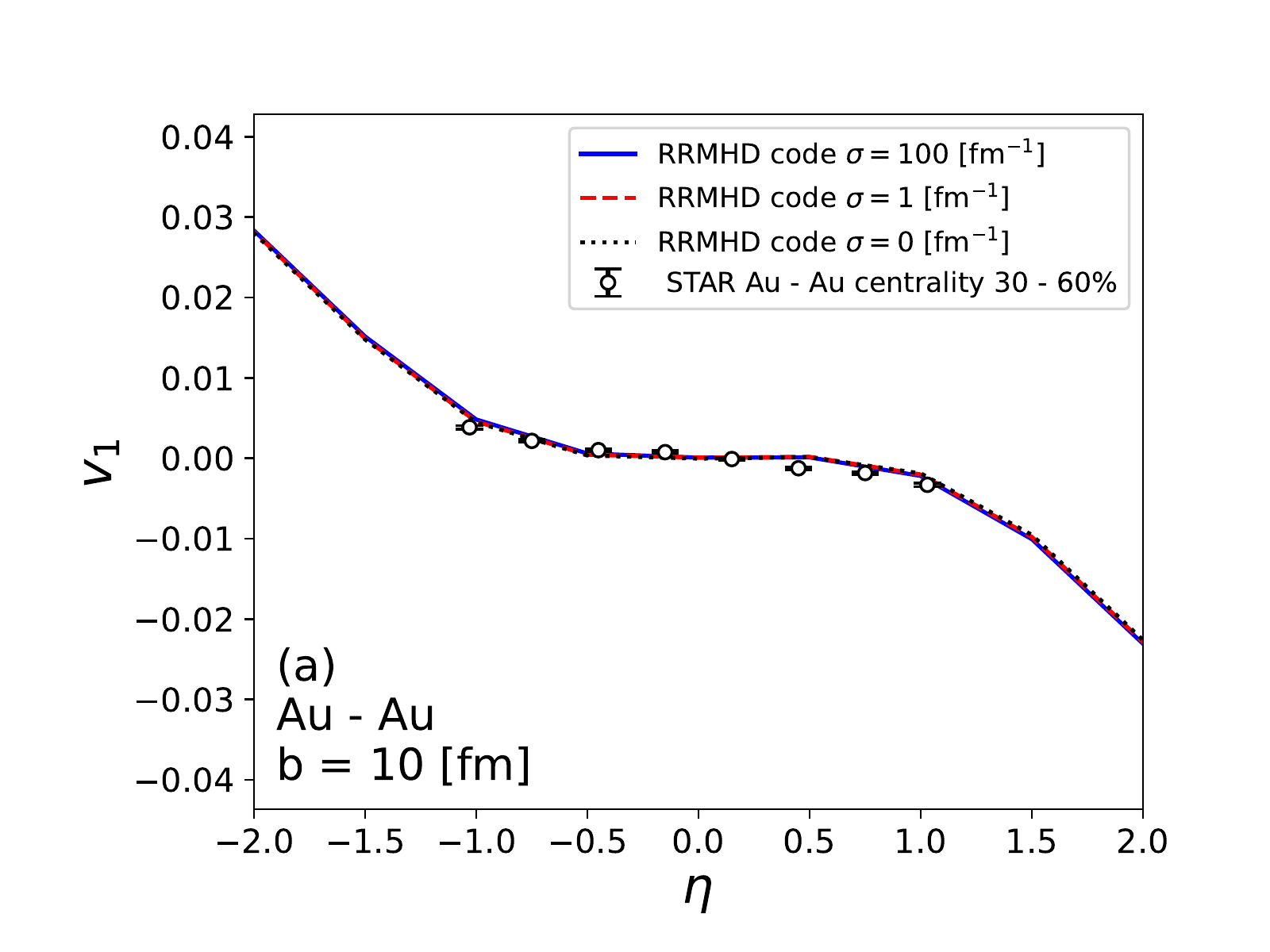}
  \end{minipage}
  \begin{minipage}[r]{0.48\linewidth}
    \includegraphics[width=9cm,height=7cm]{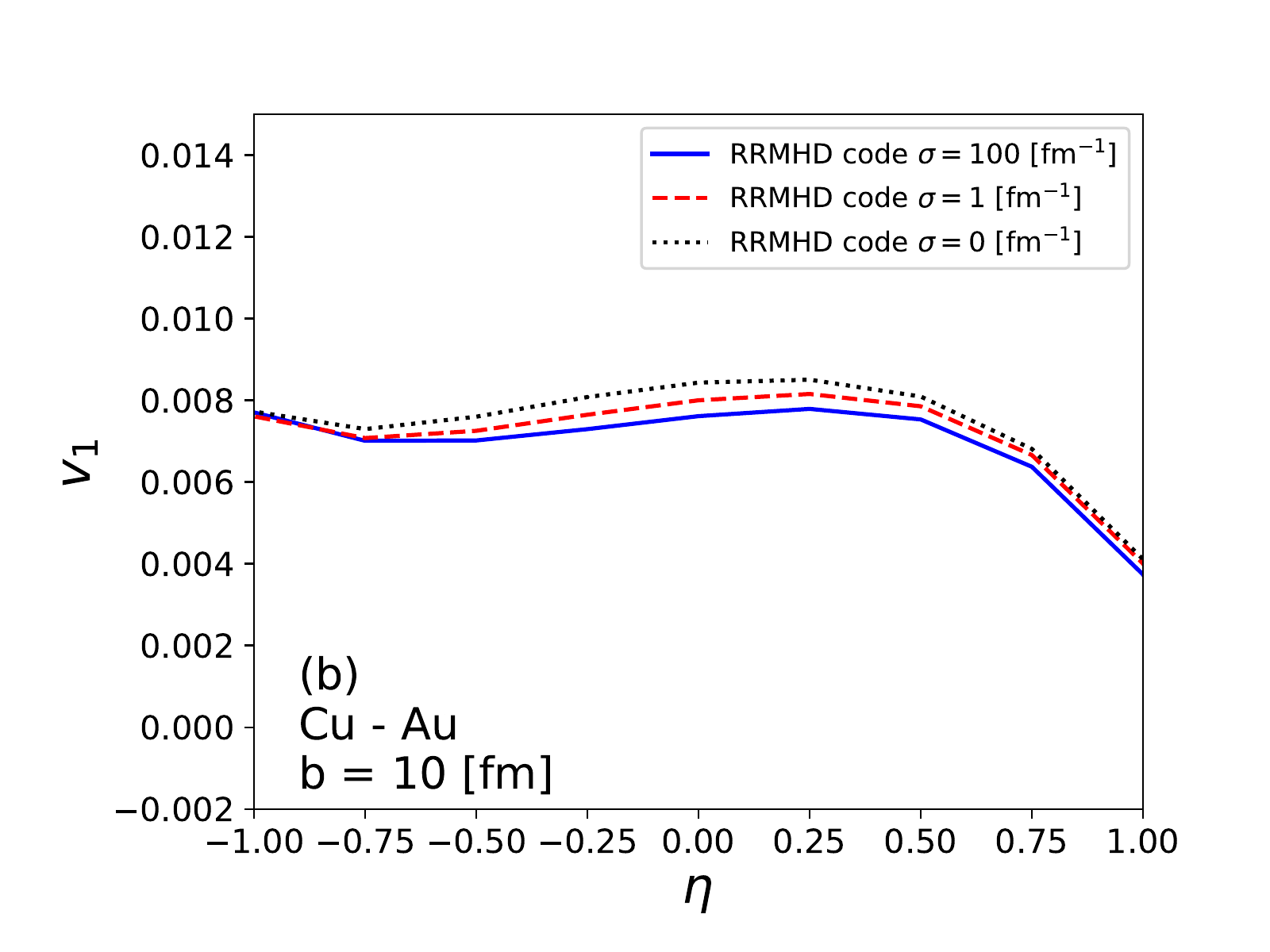}
     \end{minipage}
    \caption{(color online) The directed flow as a function of rapidity for different electrical conductivities $\sigma$. The blue solid, red dashed, black dotted lines show $\sigma = 100, 1$ and $0~\rm{fm^{-1}}$.
    We display the cases of Au-Au collisions (a) and Cu-Au collisions (b) at $\sqrt{s_{\mathrm{NN}}} = 200~\mathrm{GeV}$.
    }
    \label{fig_hydro:v1}
\end{figure*}
We investigate the effect of electromagnetic fields on the observables.
We focus on the directed flow of hadrons,
\begin{equation}
  v_1(\eta) = \frac{\int dp_\mathrm{T}d\phi \cos(\phi)\frac{dN}{dp_\mathrm{T}d\phi}}{\int dp_\mathrm{T}d\phi \frac{dN}{dp_\mathrm{T}d\phi}},
\end{equation}
where $p_\mathrm{T} =\sqrt{p_x^2+p_y^2}$ and $\phi$ is a transverse momentum and an azimuthal angle with respect to the transverse plane, respectively.
We terminate the hydrodynamic expansion at $e = 0.15~\mathrm{GeV/fm^{3}}$. 
To extract the purely hydrodynamic response of electromagnetic fields, we neglect the final state interactions. 
We adopt the Cooper-Frye formula~\cite{Cooper:1974mv} for calculation of the hadron distribution from the freezeout hypersurface.

Figure~\ref{fig_hydro:v1} shows the directed flow for the charged $\pi$ in Au-Au and Cu-Au collisions. The blue solid, red dashed and black dotted lines show the cases of the $\sigma = 100, 1 $ and $0$ $\mathrm{fm}^{-1}$, respectively.
In Fig.~\ref{fig_hydro:v1} (a), our results of the directed flow in Au-Au collisions are consistent with the STAR data in 30 - 60 \% centrality class~\cite{STAR:2008jgm}.
The electrical conductivity dependence of the directed flow is not clearly observed.
Our calculation in the high conductive case is equivalent to the ECHO-QGP simulations with the magnetic field.
In the zero conductivity case, the fluid and electromagnetic fields evolve independently.
The directed flow with the zero conductivity case is consistent with that in the relativistic ideal hydrodynamic calculation~\cite{PhysRevC.81.054902} and that without electromagnetic fields in ECHO-QGP simulations~\cite{Inghirami:2019mkc}.

We show the directed flow for charged $\pi$ in Cu-Au collisions in Fig.~\ref{fig_hydro:v1} (b).
The directed flow of our RRMHD simulation exhibits the clear dependence of the electrical conductivity of the QGP.
The amplitude of the directed flow decreases with the electrical conductivity.
This is a consequence of the reduction of the velocity in Fig.~\ref{fig_hydro:vx} (b).
In other words, the mechanism of the reduction of the directed flow is the same as the suppression of the velocity by the energy transfer from the electric field to the fluid as shown in the dissipation measure. 

We comment on the parameters of the initial condition of the medium in Cu-Au collisions.
To perform the simple comparison between the symmetric and asymmetric collision systems, we employ the same parameters of the initial condition as shown in Tab.~\ref{tab:param}.
The parameter $e_0$ which is the energy density at the $\eta_s = 0$ and $\mathbf{x}_\mathrm{T} = \mathbf{0}$ is larger than that expected in the realistic simulation for Cu-Au collisions.
Also, the plasma beta in our simulation is larger than that in the realistic case.
This means that effects of electromagnetic fields and roles of dissipation in asymmetric collision system are underestimated in this calculation.
Moreover, since we assume the constant electrical conductivity in the initial electromagnetic fields, the intensity of initial magnetic fields at the collision time is smaller than that estimated in vacuum, $|eB_y| \sim 3m^2_\pi $ at RHIC energy~\cite{MCLERRAN2014184,PhysRevC.89.054905,SUN2021136271,PhysRevC.105.054907}.
For the construction of the realistic initial electromagnetic fields, we need to solve the early dynamics of QCD matter and electromagnetic fields produced by the colliding nuclei.
We expect to observe the larger dependence of the electrical conductivity of the QGP with realistic parameter sets.
Even though that, we show that the impact of electromagnetic response on the directed flow is the same order of the viscous effect~\cite{BOZEK2012287}.
Thus, the electromagnetic field and the strength of the dissipation are important to understand phenomena in the high-energy heavy-ion collision.

The amplitude of the directed flow in our calculation is larger than the data of STAR experiment. 
Also, in $-0.75 <  \eta <  0.5$, our $v_1$ slightly increases with $\eta$, which 
is opposite tendency of $v_1(\eta)$ in the experimental data~\cite{STAR:2017ykf}.
One of the reasons for the larger value is that our calculation neglects the viscous effects and the final state interactions.
However, the viscous effect itself may not be enough to reduce the amplitude of the directed flow to the STAR data~\cite{Bozek:2012hy}. 
As a result, the directed flow in RRMHD simulation with finite viscosity may get close to the experimental data.
The final state interactions such as the hadron scattering and the resonance decay may smear the hydrodynamic response in the hadron distributions.
For the opposite tendency of $v_1(\eta)$, we may find the reason in our parameter choice for $\eta_m$ and $\eta_{\mathrm{flat}}$ which determine the rapidity profile of initial energy density and govern behavior of the directed flow in the rapidity direction.
For simplicity, we set them to be the same values in Au-Au collisions.
For the quantitative comparison with the STAR data, we need to adjust the parameters more carefully.
These remain subjects for our future works.
We conclude that the effects of the electromagnetic fields in asymmetric collision systems are sizable enough to be extracted from the experimental data.

\section{Summary}\label{V}

We have studied the consequences of the electromagnetic fields produced by the two colliding nuclei in the symmetric and asymmetric collision systems, using the RRMHD framework.
In this paper, we constructed the RRMHD model for high-energy heavy-ion collisions based on the newly developed RRMHD simulation code and investigated the directed flow in symmetric and asymmetric collision systems.


Initial conditions for the RRMHD equations are built up with the optical Glauber models~\cite{Glauber}.
In the longitudinal direction, we smoothly connected the energy density distributions from the central rapidity region to the forward and backward rapidity regions with tilted sources of the directed flow~\cite{PhysRevC.81.054902}.
For the realistic initial condition for electromagnetic fields, we have considered the solutions of Maxwell equations with the source term of the point charged particles moving in the direction of the beam axis and constant electrical conductivity of the medium~\cite{PhysRevC.88.024911}.
The parameters of the initial condition in Au-Au collisions are taken from ECHO-QGP simulation~\cite{Inghirami:2019mkc}, except for $\eta_m =3.36$~\cite{PhysRevC.81.054902}.
We employed the same parameters in Cu-Au collisions to extract the purely difference of the nucleon and charge distribution between symmetric and asymmetric collision systems.
We ignored the viscous effect and the final state interactions in order to make it clear the RRMHD response to observables. 

We found that the evolution of velocity is sensitive to electromagnetic fields in asymmetric collisions.
The electric current is induced by the electric field produced by the two different colliding nuclei because of the Ohm's law with finite electrical conductivity.
We introduced the dissipation measure~\cite{PhysRevLett.106.195003} and confirmed that a certain amount of energy transfer from the electric field to the fluid occurs in the asymmetric collision system which is not clearly observed in the symmetric collision system.

A sizable reduction of the directed flow by the dissipation has been observed.
Its magnitude is the same order of the viscous effect~\cite{Bozek:2012hy} in the asymmetric collision at RHIC energy.
The viscous effect itself may not be enough to reduce the amplitude of the directed flow to the STAR data~\cite{Bozek:2012hy}. 
The directed flow in the RRMHD with finite viscosity may get close to the experimental results. 
For quantitative analysis, we need to determine the initial parameters in Cu-Au collisions more carefully and introduce the viscous effect and the final state interactions in our model.
The directed flow decreases in the presence of the energy transfer from the electromagnetic fields to the medium.
We conclude that the effects of electromagnetic fields and the electrical conductivity in asymmetric collision systems are large enough to be detected from analysis of the experimental data.

This work can be extended in several directions.
In the presence of ultraintense electromagnetic fields, the electric charge density is induced.
It affects the charge dependent flow of hadrons~\cite{Hirono:2012rt}.
One of the promising probes of electromagnetic fields produced in high-energy heavy-ion collisions is the charge dependent azimuthal flow~\cite{PhysRevC.105.054907}.
It is possible to determine the initial electromagnetic fields.
Another direction is extension of our model to the anomalous hydrodynamics.
In heavy-ion collisions, the chiral charge density may fluctuate by the initial strong color fields. 
It induces the electric current along with the magnetic field, which is discussed as the chiral magnetic effect~\cite{KHARZEEV2008227,PhysRevD.78.074033}.
The chiral magnetic effect leads to the separation of electrical charge density.
Within the anomalous hydrodynamic calculation, the chiral magnetic effect can be observed in azimuthal anisotropy of charged hadrons~\cite{PhysRevLett.107.052303}.
The initial electromagnetic fields with the chiral magnetic effect are also investigated in Refs.~\cite{PhysRevC.91.064902,PhysRevC.105.054909}.
These theoretical predictions will give us more stringent constraints of the evolution of initial electromagnetic fields and bulk properties of QGP. 

\section*{acknowledgement}
We would like to thank Azumi Sakai for fruitful discussions.
The work of K.N. was supported in part by JSPS Grant-in-Aid for JSPS Research Fellow No.~JP21J13651.
This work was also supported by JSPS KAKENHI Grant Numbers, JP19H01928, JP20K11851 (T.M.), 
JP20H00156, JP20H11581, JP17K05438 (C.N.), JP20H00156, JP20H01941, JP20K11851, and JP21H04488 (H.R.T.).

\bibliography{v1_rrmhd} 
\bibliographystyle{unsrt} 

\end{document}